\documentstyle[12pt]{article}

\def\'#1{{\accent19\ifx #1i \i\else #1\fi}}

\def\be{\begin{equation}}
\def\ee{\end{equation}}
\def\bea{\begin{eqnarray}}
\def\eea{\end{eqnarray}}

\newcommand{\boldmathgamma}{\mbox{\boldmath$\gamma$\unboldmath}}

\newcommand{\boldmathsigma}{\mbox{\boldmath$\sigma$\unboldmath}}

\newcommand{\bfu}{{  u}}
\newcommand{\bfwt}{{\tilde{  w}}}
\newcommand{\bfw}{{  w}}

\newcommand{\bfut}{{\tilde{  u}}}

\newcommand{\bafvt}{{\bar{  V}}}
\newcommand{\bafut}{{\bar{  U}}}

\newbox\Ancha
\catcode`@=11
\newdimen\ex@
\title{Gauge and  space-time symmetry unification}

\author{J. Besprosvany\footnote{Instituto de F\'{\i}sica, Universidad Nacional Aut\'onoma de M\'exico,
Apartado Postal 20-364, M\'exico  01000, D. F., M\'exico}
}
\date{}

\begin{document}

\maketitle

%\vfill\eject

%\vspace{1in}

%\vskip30pt

%leftline{Short title:}

%\vskip40pt

%\centerline{PACS: 02.20.Qs, 03.65.Fd, 03.65.Pm}

%\vfill\eject

%\vfill\eject

\jot = 1.5ex
\def\baselinestretch{1.10}
\parskip 5pt plus 1pt

\begin{abstract}
%137 palabras!
Unification ideas suggest
an integral  treatment of fermion and boson spin and gauge-group degrees of freedom.
Hence, a   generalized quantum
field equation, based on   Dirac's, is proposed and investigated which
contains gauge and flavor
symmetries,
determines  vector
gauge field  and fermion  solution representations,
and  fixes their mode of interaction. The simplest extension of the theory
with  a $6$-dimensional Clifford algebra  predicts
an  $SU(2)_L\times U(1)$  symmetry, which is associated with the
isospin and the hypercharge, their vector carriers,
two-flavor charged and chargeless leptons, and scalar particles. %QUALIFY THIS STATEMENT ACCORDINGLY
A mass term produces  breaking of the symmetry to an electromagnetic $U(1)$, and
a Weinberg's angle  $\theta_W$
 with  $sin^2(\theta_W)=.25$\ .
  A more realistic 8-$d$ extension gives
coupling constants of the respective groups
$g=1/\sqrt{2}\approx .707$ and
$g^\prime=1/\sqrt{6}\approx .408$,  with the same $\theta_W$.

% which leads to a %Further generalization
%is shown to predict the  $SU(3)\times SU(2)_L\times U(1)$ symmetry and a three
%flavor symmetry.
\end{abstract}
\vskip 1cm

\centerline{PACS: 12.10.Dm, 03.70.+k, 12.60.-z}
\vskip 1cm

\centerline{Keywords: unification, electroweak, gauge,  couplings, leptons}

\baselineskip 22pt
\vfil\eject

\section{Introduction}

Unification has proved to  be a powerful assumption  leading to
new connections among previously considered unrelated phenomena.
It is not an exaggeration to say that most substantial advances in
the history of physics have been accompanied by the realization of
links among facts originally appearing to be independent. The
application of unification ideas differs though from one case to
the other in the scope, methods, and  results, and it is
therefore difficult to characterize it uniquely by a single rule.
Thus, the unification of known facts has sometimes led to the
prediction of  new phenomena, and these  connections have been
either experimental, theoretical, or both. It shall be useful
instead to briefly review some highlights.

The concept  of unification  is linked to the very idea of science
(or then philosophy) conceived by  the early Greek philosophers, who
 in their research of nature sought   unifying principles, although those they found were premature in their applicability.
However, a perdurable idea from those times representing  probably the most  powerful tool in physics is that
of assuming a  mathematical structure behind physical phenomena, an idea ascribed to  Pythagoras.
In modern times Galileo helped to revive the  idea of the universality of physical law in the cosmos
 by presenting supporting evidence  (e.g. the shadows
provoked  by the sun in the moon).  The principle of relativity
is a related idea he discovered which assumes this universality
for different inertial frames,
 putting powerful constraints on the possible allowed laws.
 Newton  showed, with his  new understanding of
gravity,  that the motion of cosmic and terrestrial  bodies obeys the same laws, thus
demonstrating  for  the first time a deep relation between phenomena in both expanses.

Electric and magnetic phenomena
were considered separated until the XIX Century. With the work of Amp\`ere
and Faraday it was found experimentally that one leads to the other  by changing
the kinematic state of the charges involved.
 Maxwell carried out the formalization of this into a series
of equations which provided a new understanding of light as one of
many possible waves with an electromagnetic origin, and traveling
at the speed of light,  a quantity that was predicted from the
equations.

In this century, Einstein's special relativity integrated Galileo's relativity principle with  Maxwell's equations' invariance
into a new framework by dethroning time from  its privileged use  and
putting it on a similar footing  to space, while the speed of light was assumed constant in all reference
frames.  From here a series of
new phenomena were predicted, as the equivalence  between
mass and energy.  These ideas were expanded
by linking gravity, matter, and space-time through general relativity (GR),
 a theory which assumes a geometrical framework. However, this was done only partially
since
in Einstein's GR equations only the side describing gravity and
space-time's geometry has this interpretation  while the other,
containing the energy-momentum tensor has not necessarily this form,
and waits for geometrization, if ever.
GR   predicts  new phenomena as black holes, while in the Newtonian weak gravitational
limit it produces
small corrections.

%KALUZA-KLEIN

Einstein attempted to  unify  gravity and  electromagnetism,
 but he was not really successful; nevertheless, in the meantime Kaluza and Klein developed
the idea of extending  GR  to more than 3+1 dimensions,   relating
an  additional dimension to a vector potential which could be
shown to describe the electromagnetic field. This feat
  has not led to  more information, the prediction of new phenomena,
or has been a option that  has been possible to  test but it
does show that it is a viable possibility.
Therefore, although  it cannot be
classified as a successful unification it  retains   the status of a useful working
hypothesis that is actually applied in theories as supergravity or superstrings.

Quantum mechanics (QM) successfully   accounted for the prevalence  of   particle and
wave  characteristics  encountered in  different experiments with the same objects, which is otherwise contradictory
in a classical framework.
This comprised a   completely new feature for the constituents
of nature,
which were previously thought to belong to separated classes presenting each kind
of behavior.
 The  introduction of Planck's constant required by QM
  gives rise, when using   Newton's gravity constant and the speed
of light,  to fundamental
values of mass, time,  and position; this  constitutes a unification in the sense that all measurable
quantities   can   be related to
 these  fundamental constants. A theory which would join together GR and QM should
certainly use these.

%DIRAC
One of the most beautiful examples of unification comes from Dirac who discovered a new type of
 equation that both satisfied the principles of special relativity and those  of QM.
Their  marriage in this new setting provided a new understanding of the spin 1/2 degree of freedom,
a variable previously postulated to account for various atomic phenomena and understood to be related  to
magnetic properties of fermions,
but with an otherwise inscrutable origin. Dirac's equation not only naturally gives rise
to this variable but also predicts the electron's magnetic moment with a relatively close accuracy.

In recent times the latest success in the application of unification ideas has been in
relating the weak and electromagnetic interactions in the Weinberg-Salam model (Glashow, 1961, Salam, 1968, Weinberg, 1967),  which  considers them
to originate  in a gauge symmetry,
 although their respective groups $SU(2)_L\times U(1)$  assume totally
different forms. Still, the theory succeeded  in predicting parameters
as the masses of the vectors carrying these interactions and the existence of neutral currents.

Many of today's puzzles in fundamental aspects  of physics are encountered in  the current theory of elementary particles
and fields, the standard model (SM), which involves many mysteries. Although it  is quite
successful in describing their behavior, its very construction
requires input determined by phenomenology but which is otherwise ad hoc, and  which consists of a  large
number of parameters. Worse, many aspects of this input still  need justification.
It is not clear  why there are three generations of leptons and quarks nor the origin
of their masses and  the latter's mixing angles. Neither is clear what is the source of
the parameters needed to  describe the Higgs particle, which   is as yet only a  mathematical
device to break the gauge symmetry and give masses to particles;  indeed, we lack
a more fundamental reason for the presence of a spin 0 particle.
We also lack information on  the origin of
the gauge groups of the fundamental interactions   $SU(3)\times SU(2)_L\times U(1)$,   the origin of their coupling
constants values, and   the reason
for the isospin force acting  only on a given chirality, which leads to parity violation (Lee, Yang, 1956).
%\cite{Yang}
However,
in this case, very interesting  connections  have been
obtained from grand unified models on both the forces and the values of coupling constants (Georgi, Glashow, 1974).
%\cite{unification}.
These models  assume a common
origin for these forces' gauge groups through the postulation of a group containing them as subgroups.
Still, the overall picture hints
at  a missing piece of information on an underlying principle. It may be
worth to return  to unification ideas for a clue. In particular,
we now concentrate on  the current concepts
of spin and   space-time symmetries, invoked by the first,  and from here we follow  a possible connection path
to  gauge symmetries.

While QM offers a common description for the
 shared properties of  bosons and fermions
  it still requires a specialized treatment for each to account
for their differences. Thus, while the space-time description
of the propagation of a fermion is similar to that of a boson
this  differs in the spin wave functions, which from quantum field theory  (QFT) are known
to have
a determinant influence on their very different collective behavior. A unified theory describing
both kind of particles  should address the question of their spin.
The only physical connection  comes along through  the vertex interaction which
 is determined uniquely by the gauge symmetry (e. g., in the electromagnetic
case).
Still, boson and fermion degrees of freedom are presently otherwise assumed  independent from
each other. In looking for a closer connection among them it is worth having
in mind that the spin 1/2 particle representation
of the Lorentz group ($SO(3,1)$)  is  more fundamental than the  vector one as the latter can be
obtained from a tensor product of the first  but not the other way around.

On another plane, the  fact  that a  particle   description
requires
 both configuration (or momentum) and spin spaces  leads
in turn to the fact that it is only a combination of both types of corresponding generators
that  allows for invariance under Lorentz transformations, which makes them  equally necessary.
 In this context, it is worth recalling the
Kaluza-Klein idea and wonder whether there exists a connection of the forces
of nature to extended spin spaces, instead of additional spatial dimensions.
In a way,  this idea underlay the attempt of Heisenberg  and Condon  to  understand the difference
between  the proton and the neutron. Having in mind
the similarity with spin, they assigned them with hindsight  a doublet structure,
 calling it isotopic spin or iso-spin, a concept that evolved into the   $SU(2)_L$ group
underlying the modern treatment of the weak interactions.

In this paper we  propose  a new field equation, based on   Dirac's,
 which allows for a unified treatment of
both boson and fermion spin degrees of freedom by making the solutions
share the same solution space and at the same time which
encompasses degrees of freedom which can be assigned to the gauge groups.
The equation  and  the surrounding formalism are developed in a  quantum mechanical relativistic framework,
but some aspects  of QFT will
be touched. We will show the dimensionality of the solution space
 restricts   both the possible solutions  and the
symmetries present, and that from these  an interaction prescription emerges naturally  among the field solutions.
  In particular, we obtain vertices
and their  coupling constants.  We analyze the simplest extension to 5+1  and we find
that an $SU(2)_L\times U(1)$  symmetry is predicted. The solutions will be
 hence  related to physical  fields.
 In Section II we study the  3+1 dimensional version of the new  equation  by
considering  its symmetries and a  set of commuting operators characterizing the solutions.
We also find  and analyze their   link to quantized fields.
%POSSIBILITY OF ?
  In Section III we present
its bosons solutions, both at the massless and massive levels and
in Section IV we study a particular reduction of the equation and
its transformations leading to  fermion  solutions too. We argue that
both versions of the field equation contain  a gauge invariance.
In section V we present some conserved currents and, through them
we find a link to   a vertex interaction between a pair of spin
1/2 particles and a boson, which is implied in the formalism.
 In Section VI we generalize the equation to six dimensions using
the 5+1 Clifford algebra and we analyze the embedded $4$-$d$
Clifford subalgebras, and corresponding symmetries. We  show that
for  one subalgebra chain an $SU(2)_L\times U(1)$ symmetry is
implied. In Section VII  we present the massless solutions and
link these symmetries to the isospin and hypercharge generators,
respectively.
 In Section VIII  we present the massive
ones. In Section IX we link these solutions to  physical fields in the SM, and obtain
the fermion-vector couplings and  coupling constants.
In Section  X we summarize this work, indicate its main results, and draw conclusions.

% CHANGE ORGANIZATION? GENERALIZED EQ.  VECTORS. FERMIONS.
\section{Generalized field equation from   Dirac formalism }
We search for a description of vectors and scalars as close as exists for fermions in order to be able to
relate both representations. We also demand that the  field equation which  provides
this description be  enclosed in a variational principle framework.
Indeed, these requirements   are achieved by generalizing
Dirac's equation and extending its multiplet content.
 At this point we concentrate  only on the free particle case and later on we
will show how interactions are implied in this formalism. Then,
 instead of assuming the Dirac operator acts on a spinor (Dirac, 1947)
%\cite{Dirac}
\begin{eqnarray}
\label {Diraceq}
( i \partial_\mu\gamma^\mu -M) \psi=0,
\end{eqnarray}
where $\psi$ is the column vector with components $  \psi_\alpha$,
 we  assume  it
acts on a $4\times 4$ matrix $\Psi$ with components $\Psi_{\alpha \beta}$ so that the equation becomes
\begin{eqnarray}
\label {Jaimeq}
( i \partial_\mu\gamma^\mu -M)\Psi ={ 0}.
\end{eqnarray}
The form of this equation implies   all symmetry operators valid for the Dirac equation in eq. \ref{Diraceq} (with
its corresponding particular cases of massless and massive cases)
 will be valid as well for it. The operators therefore satisfy
the Poincar\'e algebra. There  are other possible
Lorentz-invariant terms that could enter eq. \ref{Jaimeq};
 further justification for the choice of the terms in this equation is related
to a gauge symmetry  described in Section IV.

 We postulate
all transformations and symmetry operations on the Dirac operator
$( i \partial_\mu\gamma^\mu -M)\rightarrow
%SHOULD WE CHANGE THE ORDER?
{U}(\it  i \partial_\mu\gamma^\mu -M)
{U^{-1}}$ induce a corresponding transformation
\begin{eqnarray}
\label  {transfo}
 \Psi \rightarrow U \Psi U^\dagger.
\end{eqnarray}
Here, the lhs $U$ is fixed by the Dirac operator transformations
but there is a liberty for the rhs term, whose choice will
shortly prove its utility. With this assumption the elements of
$\Psi $, which can be expanded in terms of the tensor product  of
two spinors $\sum_{i,j} a_{ij} |w_i \rangle\langle w_j |$,
are expected to Lorentz transform as scalars, vectors and
antisymmetric tensors. We  will show below modified symmetry
operators classify some  solutions as fermions too.
%THIS SEEMED TO BE WRONG
%The equivalence to eq. \ref{Jaimeq} follows from the assumption that
%the action of the operator $\gamma_0( i \partial_\mu\gamma^\mu +M)$ on the right hand side
%is null;

 The vector space spanned by the matrix solutions  allows to define  an algebra to which  they belong and
which is closed. By using the matrix product, if ${ A}$, ${ B}$ are
solutions, the new field
\begin{eqnarray}
\label{field algebra}
{ C}={ AB},
\end{eqnarray}
is another element of the algebra which may  or may not  be a
solution, but lives in the same vector space. We find here a
connection to QFT as we have  an  algebra of operator solutions.
In fact, we will show the product  among fields leads to
interactions among them.

The  quantum mechanical dot product of ${ A}$, ${ B}$ is defined  by
\begin{eqnarray}
\label{pointproduct}
   \langle  { A} | { B} \rangle =tr({ A^\dagger B}).
\end{eqnarray}
A trace of over the coordinates is also implied. This definition
satisfies the usual properties expected for a measure.
% and will be used                                  DON'T SEE THE CONNECTION
%to express  zero components of
%conserved currents whose form is given below.
The use of the  product in eq. \ref{field algebra} implies the
number of terms entering the point product is not restrained  and
it may include more than two fields to be evaluated. Expectation
values of operators or any matrix element with the overlap of two
solutions can   therefore  be defined from here. An
interpretation of these products requires also taking care of the
Lorentz structure.

We note  transformation \ref{transfo} is  also valid for hermitian conjugated fields $\Psi^\dagger$,
which satisfy the equation
\begin{eqnarray}
\label {Jaimeqdagger}
{ 0}={ \Psi}^\dagger ( -i \stackrel{\leftarrow}{\partial_\mu}{\gamma^\mu}^\dagger -M).
\end{eqnarray}
We will  extend our space of solutions by
considering also combinations  of  fields ${ A}$, ${ B}^\dagger$,
  \begin{eqnarray}
\label{campechana}
 { A}+{ B}^\dagger,
\end{eqnarray}
respectively satisfying eqs. \ref{Jaimeq} and \ref{Jaimeqdagger}.
In fact, it is by taking account also  of these fields, that we can span the function space on
the  32-dimensional
 complex $4 \times 4$ matrices.

%NOT NEEDED FOR THE TIME BEING
% Although it is possible to translate this description into a form where $ M$ is
%written as
%a column matrix by using   higher dimensional Dirac matrices we will not follow this
%approach here.
{\bf Conserved Operators}

We shall be interested in plane-wave solutions of the form
\begin{eqnarray}
\label {planewaves}
{ \Psi}^{(+)}_k(x)&=&{ u}(k)e^{-i kx}\\
{ \Psi}^{(-)}_k(x)&=& { v}(k)e^{i kx}, \label{planewavesend}
 \end{eqnarray}
where $k^\mu$ is the momentum  four-vector $(E,\bf k)$, $k_0=E$.

By putting eq.  \ref{Jaimeq} into   Hamiltonian form and using the  plane-wave  states
of eqs. \ref{planewaves}, \ref{planewavesend}
each spinor
 satisfies respectively the stationary equation
\begin{eqnarray}
\label {statioJau}
\gamma_0(   {\bf k }\cdot{\boldmathgamma}+ M){ u}(k)=E { u}(k)
\end{eqnarray}
 and
\begin{eqnarray}
\label {statioJav}
\gamma_0( -  {\bf k }\cdot{\boldmathgamma}+ M){ v}(k)=-E { v}(k),
\end{eqnarray}
 with ${\boldmathgamma}=(\gamma^1,\gamma^2,\gamma^3)$.
To classify the solutions we use the Hamiltonian
\begin{eqnarray}
\label {Hamilto}
H=\gamma_0(  {\bf k }\cdot{\boldmathgamma}+ M) \end{eqnarray}
and the Pauli-Lubansky vector
\begin{eqnarray}
\label {PauliL}
 W_\mu=-\frac{1}{2}\epsilon_{\mu\nu\rho\sigma}J^{\nu\rho}p^\sigma,
\end{eqnarray}
constructed from  the Lorentz-transformation generators
\begin{eqnarray}
\label {angular}
 J_{\mu\nu}=i(x_\mu\partial_\nu-x_\nu\partial_\mu)+\frac{1}{2}\sigma_{\mu\nu},
\end{eqnarray}with the spin operators given by
 \begin{eqnarray}
 \label{sigma}
\sigma_{\mu\nu}=\frac{i}{2}[\gamma_\mu,\gamma_\nu],
 \end{eqnarray} and   momentum operator
\begin{eqnarray}
\label {momentum}
%% p^\mu=i \stackrel{\leftrightarrow}{\partial^\mu} .
 p^\mu=i {\partial^\mu}.
\end{eqnarray}
$W_\mu$  is projected over the  space-like four-vector $n_k$, orthogonal to the momentum,
  of norm $-1$ (the conventions for the norm $g_{\mu\nu}$ are given in the appendix)
\begin{eqnarray}
\label {nk}
  n_k=( \frac{{ | \bf k}| }{M}, \frac{ E {\bf k} }{M | {\bf k} |   } ),
\end{eqnarray}
giving
\begin{eqnarray}
\label {projeW} \frac{1}{M}W\cdot n_k={\bf \Sigma}\cdot{\bf \hat
k  },
\end{eqnarray}
where
\begin{eqnarray}
\label {Sigma}
{\bf \Sigma}=\frac{1}{2}\gamma_5\gamma_0{\boldmathgamma}.
\end{eqnarray}
The definition in eq. \ref{projeW} is valid both for the massless and the massive cases.

\noindent {\bf Solutions as  quantized fields}

Consistency of the definition \ref{transfo}
when applied to the generator of time translations, the Hamiltonian, implies formally that
the energy  should be obtained by taking the commutator
\begin{eqnarray}
\label {Jaimeqcomm} [\gamma_0( -i {\bf
\nabla}\cdot{\boldmathgamma}+M),\Psi ]
\end{eqnarray}
This operation calls for a rule on the action of the derivative on the right.
 We will proceed heuristically here and apply the transformation rule $p^\mu\rightarrow k^\mu$ on
 $H$.
We apply the same rule on the  $\frac{1}{M}W\cdot n_k$  operator,
which leads to ${\bf \Sigma}\cdot{\bf \hat k  }$. This
prescription is already taken into account  in   eqs.  \ref
{Hamilto} and \ref{projeW} and is as expected for spin-derivative
operators acting on a tensor product space.  We shall use this
assignment for these operators which classify the solutions
throughout this paper.
 As a bonus, we obtain that hermitian conjugates of negative energy solutions have positive energies
with the opposite spin $s$, just as occurs  in QFT, which in turn
reproduces hole theory. Indeed, assuming for the $v(k)$
component of
${ \Psi}^{(-)}_k(x)$ in eq. \ref{planewavesend},
\begin{eqnarray}
 [H   ,v(k) ]=-E  v(k)\\ \label{negenr}
[-{\bf \Sigma}\cdot {\bf {\hat k}  } ,v(k) ]= s  v(k)
\end{eqnarray}
 we find that for the hermitian conjugate wave function field
$v^\dagger(k)e^{-ikx}$, satisfying eq. \ref{Jaimeqdagger},
\begin{eqnarray}
 [H   ,v^\dagger(k)]= E  v^\dagger(k) \\ \label{holes}
[-{\bf \Sigma}\cdot{\bf {\hat k } } , v^\dagger(k) ]=- s  v^\dagger(k).
\end{eqnarray}
 We expect a more formal
justification of this operation will be given in the rigorous context of  QFT.
In addition,
consistency with eq. \ref{Jaimeq} will require  a choice of the normalization for agreement
with the energy $E$.
%However,   in both cases a non-vanishing value of $W\cdot n_k$  when
%acting on the right of  $\Psi$ requires the substitution  from the operator to the quantum number
 % ; we shall as well  invoke
%heuristically
%arguments based on this description.
% Then, we assume all operators considered here
%act in the context of a many-body Hilbert space with the no-particle one representing
%the vacuum.
% We shall assume one-body operators act on a
% homogeneous vacuum  which implies the second part of the commutator vanishes
%for the derivative part. This implies eq. \ref{Jaimeqcomm} gives the same results
%as \ref{Jaimeq} for the massless case. For the massive case the two equations  will differ due to
%the mass term.
 %When using this prescription yet for $H$ its operation on $\Psi$ is modified.

\section{Vectors, scalars, and antisymmetric tensors }

\noindent {\bf Massless solutions}

The massless equation
\begin{eqnarray}
\label {Jaimeqmle}
 i \partial_\mu\gamma^\mu  {\Psi}= { 0}
\end{eqnarray}
leads to the expressions for the  operators in   eqs.  \ref
{Hamilto} and \ref{projeW}, assuming (from here and for all
massless solutions, except when otherwise stated) that the space
component of  momentum $k^\mu$ is along the  $\hat{ \bf z}$
direction,
\begin{eqnarray}
\label {defop}
 {\bf \Sigma\cdot{\hat{\bf  k} }} =\frac{i}{2} \gamma_1\gamma_2 , \\ \label{defoHmls}
H/k_0=\gamma_0\gamma^3,
\end{eqnarray}
where the former is the helicity operator and latter is the Hamiltonian   divided by the energy.

The polarization components of the
 solutions of eq. \ref{Jaimeqmle}, bilinear in the
$\gamma$s, are given  on Tables \ref{VmA}, \ref{VpA}.
 We set the coordinate dependence as
\begin{eqnarray}
\label {coordep}
{ \Psi}^{(+)V-A }_{ki}(x)&=&{ u}_i({k} )e^{-i kx}\\
{ \Psi}^{(+)V+A }_{ki}(x)&=&\tilde { u}_i({k} )e^{-i kx}.
\end{eqnarray}
They are given together
with their quantum numbers corresponding to the operators in eqs. \ref{defop}, \ref{defoHmls}.
The solutions are also eigenfunctions of these operators $O$ in the simple form $O{ u}_i({k} )= \lambda { u}_i({k} )$
and we present the eigenvalues $\lambda$ too,
where here and throughout  the solutions are normalized  as, e. g.,
\begin{eqnarray}
\label{normalization}
tr( \bfut ^\dagger_  {i}( {k)} \bfut_{i}( {k)}) =1.
\end{eqnarray}
%e^{i \bf  p\cdot\hat {\bf x}}   dependence out
% put tilde 1 or -1 ?
%FALTA numero i en igualdad
%SEE THAT HERE NEITHER v, u are boldfaced, but on the other hand, argument is, unlike above definition. CORRECT
\begin{table}[h]
%SEE THAT HERE NEITHER v, u are boldfaced, but on the other hand, argument is, unlike above definition. CORRECT
\begin{eqnarray}
\label{lefthand} \nonumber
Vector\ solutions & \gamma_0\gamma^3 & \frac{i}{2} \gamma_1\gamma_2 \ \ [ H/k_0,\ ] \ \ [{\bf \Sigma\cdot{\hat{\bf  p} }} ,\ ] \\ \nonumber
\bfu_{-1}({k}) =\frac{1}{4} (1-\gamma_5)\gamma_0(\gamma_1-i \gamma_2)& 1 & -1/2\ \ \ 2\ \ -1\\ \nonumber
\bfu_{-1}({ \tilde{k}}) =\frac{1}{4} (1-\gamma_5)\gamma_0(\gamma_1+i \gamma_2)&  -1 & 1/2\ \ 2 \ \ -1\\ \nonumber
\bfu_0( {k})= \frac{1}{4} (1-\gamma_5)\gamma_0(\gamma_0-\gamma_3)& 1 & -1/2\  \ \ \ 0 \ \ \ 0 \\ \nonumber
\bfu_{0}( {\tilde  k)} =\frac{1}{4} (1-\gamma_5)\gamma_0(\gamma_0+\gamma_3)& -1 & 1/2 \ \ \ 0\ \ \ 0   \nonumber
\end{eqnarray}
\caption{\label{VmA}V-A\   terms. }
\end{table}
\begin{table}
\begin{eqnarray}\nonumber
Vector\ solutions & \gamma_0\gamma^3 & \frac{i}{2} \gamma_1\gamma_2 \ \ [ H/k_0,\ ] \ \ [{\bf \Sigma\cdot{\hat{\bf  p} }} ,\ ] \\ \nonumber
\bfut_{1}( {k} ) =\frac{1}{4} (1+\gamma_5)\gamma_0(\gamma_1+i \gamma_2)
 & 1 & 1/2  \ 2 \ \ 1\\ \nonumber
\bfut_{1} ({ {\tilde k}})=  \frac{1}{4} (1+\gamma_5)\gamma_0(\gamma_1-i \gamma_2)
& -1 & -1/2\  \ \ 2  \ \ \ 1\\ \nonumber
\bfut_{0}( {k)} =\frac{1}{4} (1+\gamma_5)\gamma_0(\gamma_0-\gamma_3)& 1 & 1/2\ \ \ 0  \ \ \ 0  \\ \nonumber
\bfut_{0}( {\tilde k)}  =\frac{1}{4} (1+\gamma_5)
\gamma_0(\gamma_0+\gamma_3)& -1 & -1/2\ \ \ 0 \ \ \ 0
    \nonumber  %CHECK QUANTUM NUMBERS
\end{eqnarray}
\caption{\label{VpA}V+A\  terms.   }
\end{table}
%NOtice last strange formula. Put other form to all or take out? Correct others
Solutions $\bfu_{-1}({k})$, $\bfut_{1}( {k} )$ correspond to on-shell particles with transverse polarizations but
opposite helicities,  while  the off-shell $\bfu_{0}({k})$, $\bfut_{0}( {k} )$
are polarized in the  longitudinal-scalar directions.
All these solutions correspond  to waves propagating  in the
$\hat {\bf z}$ direction.  The other terms propagate in the
  $-\hat {\bf z}$ direction, which is represented by  four-vector  $\tilde  k^\mu=k_\mu$,
  and are classified with the appropriate  relations as eqs. \ref{defop} and \ref{defoHmls}.

The coordinate dependence of these solutions is given by
\begin{eqnarray}
\label {coordeptil}
{ \Psi}^{(+)}_{\tilde k i}(x)&=&{ \bfu}_i({\tilde k} )e^{-i  \tilde k x }.
 \end{eqnarray}

These solutions  do not represent independent polarization components as e.g.  $\bfu_{i}({\tilde k})$
can be obtained by rotating the $\bfu_{i}({k})$.
 The classification $V$+$A$ and $V$-$A$,   consisting respectively of the   $\bfut _{i}$ and $\bfu _{i}$ terms,
 corresponds to specifying the weight of vector and axial components,
which is further clarified
 below. These two
types of  solutions  are also characterized by
the two vector spaces projected  by $\frac{1}{2} (1+\gamma_5)$ and $\frac{1}{2} (1-\gamma_5)$
which they generate respectively but which they do not exhaust.
We need to consider the negative energy solutions
\begin{eqnarray}
\label {coordepv}
{ \Psi}^{(-)}_{ki}(x)&=& { v}_i({k} )e^{i kx}
\end{eqnarray}
and  use their hermitian conjugates, which
in fact generate other polarization components, in order to   completely   span the space.
In the massless case we have negative energy solutions ${ v}_{ i}({k} )=\bfu_{i}(k)$,
 $\tilde { v}_{ i}({k} )=\bfut_{i}(k)$ (and $\tilde k$ terms), that is, with opposite helicities.
The  combinations of the type \ref{campechana}
$\frac{1}{\sqrt{2}}[\tilde { u }_i({k})\pm\tilde { v }_{ i} ^\dagger(k)] e^{-i    kx}$,
 $\frac{1}{\sqrt{2}}[{ u }_i({k})\pm { v }_{  i}^\dagger (k)  ] e^{-i    kx} $ $({ v }_{  i}^\dagger (k)\equiv[{ v }_{  i}(k)]^\dagger )$
   will be interpreted
as vector solutions with varied polarizations.
The chirality operator $\gamma_5$ further characterizes these solutions as non-chiral since,
 using  rule \ref{transfo},
it gives $[\gamma_5 ,\Psi ]=0$.
The most general form of the solutions can be obtained by rotating and boosting these solutions through
a Lorentz transformation, using $J_{\mu\nu }$ in eq. \ref{angular}.

Eq. \ref{Jaimeq}  also satisfies  the discrete invariances of time,  and  space inversion, and charge conjugation,
expressed  respectively by the operators
\begin{eqnarray}
\label{CPT}
T&=&i\gamma_1\gamma_3 \mathcal{K} \mathcal{T}   \label{T} \\
P&=&\gamma_0 \wp  \label{P} \\
C&=&i\gamma_2 \mathcal{K} \label{C},
\end{eqnarray}
where $\mathcal{K}$ is the complex conjugation
operator $\mathcal{K}$$ i $$\mathcal{K}=$$-i$,  $\mathcal{T}$ changes $t\rightarrow -t$, and $\wp$
 changes ${\bf x}\rightarrow -{\bf x}$
and  consequently ${\bf p}\rightarrow -{\bf p}$;  we use the Dirac representation for the $\gamma_\mu$ matrices
(see appendix).
It is then possible to form combinations of the above solutions transforming as vectors and as axial vectors.
For example, the combination  \begin{eqnarray}
 \label{goodparity}
  { \Psi} _{k {\hat x} }  = \frac{i}{2}[ \tilde { u }_1({k})+{ u }_{-1}({k})+
\tilde { v }_{1}^ \dagger (k) + { v }_{-1}^\dagger (k)
  ] e^{-i    kx} =\frac{i}{2}\gamma_0\gamma_1 e^{-i    kx}
 \end{eqnarray}
represents  a vector particle linearly polarized along
$\hat {\bf x}$, that is, it transforms  into $-{ \Psi}_{k{\hat x} } ({\tilde x})$ under $P$, with $\tilde  x_\mu=x^\mu$. In
general
 \begin{eqnarray}
\label{Amufirst}
{ A}_\mu(x) =\frac{i}{2}\gamma_0\gamma_\mu e^{-i    kx}
\end{eqnarray}
(and the corresponding negative energy solution)  transforms  into ${ A}^\mu({\tilde x})$ under $P$,
into ${ A}^\mu(-{\tilde x})$ under $T$,  and
into $-{ A}_\mu({-x})$ under $C$.
 \begin{eqnarray}
\label{AmufirstAx}
{ A} _ {5\mu} (x) =\frac{i}{2}\gamma_5\gamma_0\gamma_\mu e^{-i    kx}
\end{eqnarray}
transforms  into $-{ A}^{\mu}_5 ({\tilde x})$ under $P$, into ${ A}_5^\mu(-{\tilde x})$ under $T$,    and
into ${ A}_{5\mu}({-x})$ under $C$.
The combination $ { A}_\mu(x)  +C { A}_\mu(x) C^\dagger$ transforms
into minus itself under charge conjugation, as expected for a non-axial vector. Given the quantum
numbers of  ${ A}_\mu(x)$  it becomes possible
to relate it  to the vector
potential of an electromagnetic field. Indeed, similar mixtures of   ${ \tilde u}$, ${\tilde  v}^\dagger$,
${ u}$, and ${ v}^\dagger$
 solutions
have been shown, under certain conditions, to satisfy
 Maxwell's equations (Bargmann, Wigner, 1948).
%\cite{Bargmann}.

The remaining eight degrees of freedom   in the massless case are classified
into six forming an
antisymmetric tensor  and two scalars, which as solutions appear
mixed. The chirality $\gamma_5$ further divides them into left and right-handed.  Their respective
coordinate dependence is
\begin{eqnarray}
\label {coordepw}
{ \Psi}^{(+)-}_{ki}(x)&=&{ w}_i({k} )e^{-i kx}\\
{ \Psi}^{(+)+}_{ k i}(x)&=&\tilde{ w}_i({ k} )e^{-i  k x } \label{coordepwtilde}
 \end{eqnarray}
(and corresponding definitions for $\tilde k$)
and the explicit form of the matrix components together with their quantum numbers
is shown on  Tables \ref{eigenantis}, \ref{eigsend}.
\begin{table}[h]
\begin{eqnarray}\nonumber
Scalars\  and\ antisymmetric\ tensors &   \gamma_0\gamma^3 & \frac{i}{2} \gamma_1\gamma_2\ \ [ H/k_0,\ ]   \ \  [{\bf \Sigma\cdot{\hat{\bf  p} }} ,\ ] \nonumber  \\
\bfw_{0} ({k})= \frac{1}{4} (1-\gamma _5) ( \gamma_0 +\gamma _3)& 1 & -1/2 \ \ 2 \ \ 0  \nonumber\\
\bfw_{0} ({ \tilde k})= \frac{1}{4} (1-\gamma _5)  ( \gamma_0 -\gamma _3)& -1 & 1/2 \ \  2 \ \ 0 \nonumber \\
\bfw_{-1}({k})= \frac{1}{4} (1-\gamma_5)(\gamma_1  -i\gamma _2)& 1 &  -1/2 \ \ 0 \ \ -1 \nonumber \\
\bfw_{-1}({\tilde k})= \frac{1}{4} (1-\gamma_5)(\gamma_1+i\gamma_2)& -1 & 1/2 \ \  0 \ \ -1  \nonumber   %NOTESE QUE
%LA TILDE VA CON EL +, ya que la chirality es NEGATIVA
\end{eqnarray}
\caption{left-handed bosons.\label{eigenantis} }
\end{table}

\begin{table}
\begin{eqnarray}\nonumber
Scalars\  and\ antisymmetric\ tensors &   \gamma_0\gamma^3 & \frac{i}{2} \gamma_1\gamma_2\ \ [ H/k_0,\ ] \nonumber  \ \  [{\bf \Sigma\cdot{\hat{\bf  p} }} ,\ ] \nonumber  \\
\bfwt_{0} ({ k})= \frac{1}{4} (1+\gamma_5)(\gamma_0+\gamma_3)& 1 & 1/2 \ \  2  \ \ 0 \nonumber\\
 \bfwt_{0} ({ \tilde  k})= \frac{1}{4} (1+\gamma_5)(\gamma_0-\gamma_3)& -1 & -1/2 \ \  2 \ \ 0  \nonumber\\ %CHECK NUMBERS
\bfwt_{1} ({ k})= \frac{1}{4} (1+\gamma_5)(\gamma_1+i\gamma_2)& 1 & 1/2 \ \  0 \ \   1  \nonumber\\
\bfwt_{1} ({ \tilde   k})=\frac{1}{4}  (1+\gamma_5)(\gamma_1-i\gamma_2)& -1 & -1/2 \ \  0 \ \   1  \nonumber
\end{eqnarray}
\caption{right-handed bosons.\label{eigsend}}
\end{table}

To see these terms have this interpretation we should apply transformation \ref{transfo}
with $U$ containing a Lorentz transformation, acting
on $1 { \Psi}=\gamma_0\gamma_0 { \Psi}$, which leads to
 $U^\dagger \gamma_0=\gamma_0 U^{-1}$.
Labeling the antisymmetric terms by
\begin{eqnarray}
\label{laanti}
{ A } _{\mu\nu}=\frac{1}{4}\gamma_ 0 [\gamma_ \mu,\gamma_ \nu],
 \end{eqnarray}
 and the scalar and pseudoscalar terms by
\begin{eqnarray}
\label{lascapse}
 {\phi}=\frac{1}{2} \gamma_0, \\
 {\phi_5}=\frac{1}{2} \gamma_0\gamma_5,
\end{eqnarray}
 the expressions  on Tables \ref{eigenantis}, \ref{eigsend}
 can be written in terms of $A_{\mu\nu}$,
 $\phi$, and $\phi_5$. This requires also hermitian
conjugates of negative energy solutions
\begin{eqnarray}
\label {coordepwneg}
{ \Psi}^{(-)-}_{ki}(x)&=&{ z}_i({k} )e^{i kx}\\
{ \Psi}^{(-)+}_{\tilde k i}(x)&=&\tilde { z}_i({ k} )e^{i k x }, \label{coordepwtildeneg}
 \end{eqnarray}
where
${ z}_{i}({k} )=\bfw_{i}(k)$,
 $\tilde { z}_{i}({k} )=\bfwt_{i}(k)$ (and $\tilde k$ terms).
While the scalar and pseudoscalar particles obtained have a straightforward interpretation as on-shell
particles the antisymmetric
solutions do not have a recognizable interpretation, given that their on-shell components do
not have transverse polarizations. A  vector interpretation  can be given using
the identities
\begin{eqnarray}
\label {identities}
 \bfwt_{i}({k}) =\frac{1}{2|{\bf k}|}\not\! k  \bfu_{-i}({\tilde k}),\ \
 \bfwt_{i}({\tilde k}) =\frac{1}{2|{\bf k}|}\not\! {\tilde k}  \bfu_{-i}({k}),  \\ \nonumber
\bfw_{-i}({k}) =\frac{1}{2|{\bf k}|}\not\!{k}  \bfut_{i}({\tilde k}),\ \
\bfw_{-i}({\tilde k}) = \frac{1}{2|{\bf k}|}\not\! \tilde k \bfut_{i}({k}),  \nonumber
\end{eqnarray}
and similar expressions for negative energy solutions. The gauge symmetry discussed
below suggests some of these  solutions may be gauged out.

\noindent {\bf Polarization vectors}

The solutions presented so far on Tables  \ref{VmA}, \ref{VpA}
and \ref{eigenantis}, \ref{eigsend} are given in terms of
components that are eigenstates of the helicity operator and are
therefore components of   spherical harmonic  vectors. In
general, we can show the solutions generate a quadrivector basis
whose  components can be given in a spherical or in  a vector
basis.

A set of corresponding polarization vectors  $\epsilon^{(\lambda)}(k)$ can be
defined which coincide with the directions  that some of the
actual solutions in  Tables  \ref{VmA}, \ref{VpA}  and
\ref{eigenantis}, \ref{eigsend} take. We define a unitary  vector
$n$, along the time direction, that is, $n^2=1$.
%$n_k$ orthogonal to $k$: $n_{k\mu}=tile k_\mu=k^\mu$. Then $k^\mu n_{k\mu}=0$.
Assuming a general $k$ we choose
$\epsilon^{(1)}(k)$, and $\epsilon^{(2)}(k)$ in the transverse directions, orthogonal
to $k$ and $n$, and   $\epsilon^{(\lambda)}(k)\cdot\epsilon^{(\lambda^\prime)}(k)=-
\delta^{\lambda\lambda^\prime}$. Then we pick $\epsilon^{(3)}(k)$, the  longitudinal vector,   along the plane
$k$-$n$ and orthogonal to $n$ and $\epsilon^{(0)}(k)$, the   scalar component,  along
$n$. These vectors
are orthogonal among themselves:
\begin{eqnarray}
\label{orthogoneps}
\epsilon^{(\lambda)}(k)\cdot\epsilon^{(\lambda^\prime)}(k)=
g^{\lambda\lambda^\prime}.
\end{eqnarray}
%put here further relations among vectors?

In the case of  the  solutions $\tilde u_i$ on Table  \ref{VpA},
which propagate along $\pm\bf {\hat z}$,
 the polarization vectors in the  spherical basis are
\begin{eqnarray}
\label{polarizsim}
e^{ (1)}(k)&=& \bfut_1({k}) , \\
e^{ (2)}(k)&=&  \bfut_1({\tilde k}) , \\
e ^{ (3)}(k)&=&\frac{1}{\sqrt{2}} (\bfut_0({\tilde k})-\bfut_0({ k})), \\
e^{ (0)}(k)&=&\frac{1}{\sqrt{2}} (\bfut_0({\tilde k}) +\bfut_0({ k}))
\end{eqnarray}
The associated vector form of the polarizations is given by
$\frac{1}{2\sqrt{2}}(1+\gamma_5)\gamma_0\gamma_\mu$. In fact, the
sixteen  components of the four vectors
 $\epsilon^{(\lambda)}(k)$ form a tensor which connect among the two bases.
%CHECK THIS WORD
The components are obtained from
\begin{eqnarray}
\label{polarizcompo}
\epsilon^{(\lambda)}_\mu=tr(e^{*(\lambda)}(k)\frac{1}{2\sqrt{2}}(1+\gamma_5)\gamma_0\gamma_\mu),
\end{eqnarray}
where we use the conjugate polarizations
\begin{eqnarray}
\label{polariz}
e^{*(1)}(k) &=&\frac{1}{4 }(1+\gamma_5)(\gamma_1-i \gamma_2)\gamma_0 \\
e^{*(2)}(k)&=&\frac{1}{4 }(1+\gamma_5)(\gamma_1+i \gamma_2)\gamma_0 \\
e^{*(3)}(k)&=&\frac{1}{2\sqrt{2}} (1+\gamma_5)\gamma_3\gamma_0\\
e^{*(0)}(k)&=&\frac{1}{2\sqrt{2}}(1+\gamma_5).
\end{eqnarray}

For the  non-axial vectors of the form \ref{goodparity} the terms  $\frac{1}{{2}}\gamma_0\gamma_\mu$ constitute the
the vector basis.
Indeed, we can use the relation
\begin{eqnarray}
\label{orthogon}
tr[(\gamma_0\gamma_\mu)(\gamma_0\gamma_\nu )]=tr[(\gamma^\mu\gamma_0)(\gamma_0\gamma_\nu )]=4 g_{\ \nu}^\mu
\end{eqnarray}
in order to project precisely those components; namely, we should seek
\begin{eqnarray}
\label{components}
C_{\Psi}^\mu=tr(\frac{1}{2}\gamma^\mu\gamma_0 \Psi)
\end{eqnarray}
%where we obtain non-zero components for solutions \ref{righthand}-\ref{lefthandend}
%check notation
($\gamma_0$ is included to account for the other
 $\gamma_0$ factor  that is  included in the solutions).

For solutions $w_i$, $\tilde w_i$ on Tables
\ref{eigenantis}, \ref{eigsend}   an orthonormal vector basis can be found in the vector interpretation of eq.
\ref{identities} which
contains them. This is obtained by using, for example, the vectors
\begin{eqnarray}
\label{orthogonoth}
b^\mu=i\gamma_0\frac{1}{2\sqrt{-\Box}}(\gamma^\mu {\not\! \partial}+ \sqrt{2}\partial^\mu ) \\
{b^\mu}^*=-i\frac{1}{2\sqrt{-\Box}}(\gamma^\mu {\not\! \partial}- \sqrt{2}\partial^\mu )\gamma_0 \nonumber
\end{eqnarray}
 which also satisfy
\begin{eqnarray}
\label{orthogongamma}
tr(b_\mu^* b_\nu )=g_{\mu\nu},
\end{eqnarray}
 as can be shown by using the relation
$tr(\gamma_\mu\not\!\partial\gamma_\nu{\not\! \partial })=4
  (2 \partial_\mu\partial_\nu-g_{\mu\nu}\Box)$.
The presence of the $\partial_\mu$ in these expressions uses the fact that
we may generate a vector solution from a scalar by taking the derivative.
Given that we have constructed solutions
that satisfy
Dirac's equation \ref{Diraceq} it follows the solutions will also
satisfy the Klein-Gordon equation. This means that
when  projecting the  solutions on vectors
\ref{orthogonoth} these can only be defined as a limiting case as
the $\frac{1}{\sqrt{-\Box}}$ operator is  singular when applied on the
solutions.

\noindent {\bf Gauge invariance}

A clue for the  interpretation  of all massless solutions described
so far is suggested by the fact that  eq. \ref{Jaimeq} is invariant to first order
under a set of gauge  transformations,  that is,
with local dependence, which implies some are spurious.
 % CHECK THIS STATEMENT These symmetries turn out to be
%trivial when considering Dirac's equation (\ref {Diraceq}), but they are not when
%applied in the sense of eq.
In trying to generate this transformation we expect it  to be unitary and Lorentz invariant.
However, we can only present a transformation satisfying
either property but not both together. However, we expect that it satisfy both properties when
applied on the space of solutions.
This is reminiscent of the QFT case.

 We now consider the  transformation $U_G=e^{iG}$ with a similar  sense
to eq. \ref{transfo}, with generator  $G={\not\! \partial a(x)}$,
and $a(x)$ any
real  function, where we use  the form $H\rightarrow  \tilde U ^\dagger H \tilde U$.
When applying the corresponding  infinitesimal transformation   to the operator $i\gamma_0 {\not\! \partial}$
we need to consider only the
 commutator (or anticommutator, if we take $i{\not\! \partial a(x)}$  as the generator)
with the Dirac operator, which contains
\begin{eqnarray}
\label{commugaugetrans}
[\not\! \partial,\not\! \partial a(x)]_\pm=
\Box a(x)
\pm \partial_\mu a(x)\partial_\nu \gamma^\mu \gamma^\nu.
\end{eqnarray}
The (anti-)commutator  with the operator $a(x)\not\! \partial$
gives
\begin{eqnarray}
\label{commugaugeoth}
[\not\! \partial ,a(x)\not\! \partial ]_\pm= \partial_\mu a(x) \partial_\nu \gamma^\mu \gamma^\nu\pm
a(x) \Box.
\end{eqnarray}
From these equations we see
 $U_{G^a}=e^{i G^a } $,
where
\begin{eqnarray}
\label{genegaugeg}
G^a= \not\!\partial a(x)+{a(x)\not\!\partial}
\end{eqnarray}
(or the transformation with generator $\not\!\partial a(x)-{a(x)\not\!\partial}$) will be invariant to first order
provided $\Box a(x)+2 \partial_\mu a(x) k^\mu=0$.
Consequently, the  symmetry is linked to the space of solutions. We also get a cancellation to
 second order in $G^a$ if $\Box a(x)=0$, and $ \partial_\mu a(x) k^\mu=0$  are satisfied.
We note
these  conditions mean
$a(x)$ satisfies the massless Klein-Gordon equation in a reduced number of directions.
 Although $U_{G^a}$ is a Lorentz-invariant operator it
is not hermitian.

The term
\begin{eqnarray}
\label{genegaugeg5}
G_5^b=i[b(x)\gamma_5\not\! \partial+ {\gamma_5 \not\! \partial b(x)}]
\end{eqnarray}
under the conditions  $\Box b(x)=0$ and  $\partial_\mu b(x) k^\mu=0$ is similarly the
generator  of
another symmetry   operator $U_{G_5^b}=e^{i G_5^b}$ since
\begin{eqnarray}
\label{commugaugeg5}
[\not\! \partial, \gamma_5 \not\! \partial b(x)]_\pm=
\gamma_5 (-\Box b(x)
\pm \partial_\mu b(x)\partial_\nu \gamma^\mu \gamma^\nu),
\end{eqnarray}
and
\begin{eqnarray}
\label{commugaugetransg5}
[\not\! \partial ,b(x)\gamma_5 \not\! \partial ]_\pm= \gamma_5(- \partial_\mu b(x) \partial_\nu \gamma^\mu \gamma^\nu\pm
b(x) \Box).
\end{eqnarray}
The  case with different $V,$ $A$ contributions is  considered below.

We have obtained   two sets of local transformations
 restrained  by the condition on  the functions $a(x)$ and $b(x)$.
  We may understand this as a manifestation of a gauge
symmetry, where we  attribute  the restriction
to  the choice  of gauge.
Indeed, we find a  similarity with the gauge invariance  of the  electromagnetic field $A_\mu$.
The Lorentz  gauge condition  (Lorentz invariant) for it
\begin{eqnarray}
\label{Lorentz}
\partial^\mu A_\mu=0
\end{eqnarray}
reduces Maxwell's equations to
\begin{eqnarray}
\label{Maxwell}
\Box A_\mu=0.
\end{eqnarray}
In this case  the gauge freedom is reduced to transformations
$A_\mu\rightarrow A_\mu+ \partial_\mu \phi$, where $\phi$
satisfies $\Box \phi=0$. The fact that solutions on Tables
\ref{VmA}, \ref{VpA} and \ref{eigenantis}, \ref{eigsend}  also
satisfy the massless Klein-Gordon equation
 supports the interpretation of these solutions as vector particles satisfying Maxwell's
equation within the Lorentz gauge.
In fact, these solutions resemble more the case of the quantized electromagnetic
field in this gauge. It is easy to see that this set of
solutions does not satisfy eq. \ref{Lorentz}. However,  eq. \ref{Jaimeq}
can be interpreted as implying that the Lorentz condition is satisfied in the mean, a condition
required for the quantized electromagnetic field
\begin{eqnarray}
\label{quanLorentz}
 \partial^\mu A_\mu|\psi\rangle =0, \end{eqnarray}
 where $|\psi\rangle $
 describes states from the electromagnetic field.
To put eq. \ref{Jaimeq} in
this form we only need to use combined solutions as obtained in eq. \ref{goodparity}.

It is interesting that in our case eq. \ref{quanLorentz} is
 a condition that we derive and not one that we impose additionally from gauge fixing.
We therefore obtain again a connection to QFT.
The suggested gauge symmetry  also would imply not all the solutions in the fields on Tables
\ref{eigenantis},\ref{eigsend} are independent, but rather that they could be obtained
from the fields on Tables  \ref{VmA}, \ref{VpA}
by a gauge transformation.
 %CHECK THIS STATEMENT IS THE GAUGE SYMMETRY NOT TRIVIAL
The presence of a gauge symmetry places constraints on the
choices of terms in a quantum relativistic equation, as happens in QFT. Thus, here we
find a justification for the choice of Lorentz-invariant terms in eq.
 \ref{Jaimeq}. %PUT IN CONCLUSIONS?

\noindent   {\bf Massive solutions}

In order to describe the solutions of eq. \ref{Jaimeq} with $M \not= 0$  we choose the rest frame so that they only
have time dependence
\begin{eqnarray}
\label {massplanewaves}
{ \Psi}^{(+)M}_{ki}(x)&=&{ U_i}(k)e^{-i Mt}\\
{ \Psi}^{(-)M}_{ki}(x)&=& { V_i}(k)e^{i Mt}. \label{massplanewavesend}
 \end{eqnarray}
The matrix components are classified by the eigenvalue of the parity operator $P$ into the  $P=-1$ group
on Table \ref{massivesol}
\begin{table}[h]
\begin{eqnarray}\nonumber
Massive\ bosons & \gamma_0 & \frac{i}{2} \gamma_1\gamma_2\ \ [ H/k_0,\ ]   \ \  [{\bf \Sigma\cdot{\hat{\bf  k} }} ,\ ] \nonumber \\
 {{ U}}_{1}(M,{\bf 0})= \frac{1}{4} (1+\gamma _0) (\gamma_1+i\gamma_2)&  1 & 1/2 \ \  2\ \ 1 \nonumber \\
{{ V}}_{1}(M,{\bf 0})= \frac{1}{4} (1-\gamma _0) (\gamma_1+i\gamma_2)& -1 & 1/2 \ \  -2 \ \ 1  \nonumber\\
{{ U}}_ {-1} (M,{\bf 0})= \frac{1}{4} (1+\gamma _0) (\gamma_1-i\gamma_2)& 1 & -1/2 \ \  2 \ \ -1 \nonumber \\
{{ V}}_ {-1} (M,{\bf 0})= \frac{1}{4} ( 1-\gamma _0) (\gamma_1-i\gamma_2)& -1 & -1/2 \ \ -2\ \ -1  \nonumber\\
{{ U}}_ {0}(M,{\bf 0})=  \frac{1}{4} (1+\gamma _0) ( \gamma_5 -\gamma _3)&  1&  1/2  \ \ 2\ \  0 \nonumber\\
{{ V}}_{0}(M,{\bf 0})=  \frac{1}{4}(1-\gamma _0)  ( \gamma_5+ \gamma _3)& -1 &  1/2  \ \ -2 \ \  0 \nonumber\\
 {{ U}}_{\tilde 0}(M,{\bf 0})= \frac{1}{4} (1+\gamma _0) ( \gamma_5 +\gamma _3)&  1 & -1/2 \ \ 0  \ \  2\ \ 0 \nonumber \\ \nonumber
 {{ V}}_{\tilde 0} (M,{\bf 0})=  \frac{1}{4}(1-\gamma _0)  ( \gamma_5- \gamma _3)& -1 & -1/2 \ \ -2 \ \  0
\label{massivesolend}
\end{eqnarray}
\caption{Parity   $P=-1$ massive bosons.\label{massivesol}}
\end{table}
and $P=1$ group on  Table \ref{massivesolPmend}
\begin{table}[h]
\begin{eqnarray}\nonumber
Massive\ bosons  & \gamma_0 & \frac{i}{2}  \gamma_1\gamma_2\ \ [ H/k_0,\ ]   \ \  [{\bf \Sigma\cdot{\hat{\bf  k} }} ,\ ] \nonumber \\
{{\bafut}}_{1}(M,{\bf 0})= \frac{1}{4} \gamma_5(1-\gamma _0) (\gamma_1+i\gamma_2)& 1 & 1/2 \ \  0\ \ 1 \nonumber \\
{{\bafvt}}_{1}(M,{\bf 0})= \frac{1}{4} \gamma_5(1+\gamma _0) (\gamma_1+i\gamma_2)& -1 & 1/2 \ \  0 \ \ 1 \nonumber \\
{{\bafut}}_ {-1} (M,{\bf 0})= \frac{1}{4} \gamma_5(1-\gamma _0) (\gamma_1-i\gamma_2)& 1 & -1/2 \ \  0 \ \ -1 \nonumber\\
 {{\bafvt}}_ {-1} (M,{\bf 0})= \frac{1}{4} \gamma_5( 1+\gamma _0) (\gamma_1-i\gamma_2)& -1 & -1/2 \ \ 0\ \ -1 \nonumber \\
{{\bafut}}_ {0}(M,{\bf 0})=  \frac{1}{4} \gamma_5(1-\gamma _0) ( \gamma_5 +\gamma _3)&  1&  1/2  \ \ 0 \ \  0 \nonumber\\
  {{\bafvt}}_{0}(M,{\bf 0})=  \frac{1}{4}\gamma_5(1+\gamma _0)  ( \gamma_5- \gamma _3)& -1 &  1/2  \ \ 0 \ \  0 \nonumber\\
{{\bafut}}_{\tilde 0}(M,{\bf 0})= \frac{1}{4}\gamma_5 (1-\gamma _0) ( \gamma_5 -\gamma _3)&  1 & -1/2 \ \ 0  \ \  0\nonumber \\ \nonumber
{{\bafvt}}_{\tilde 0}(M,{\bf 0})=  \frac{1}{4}\gamma_5(1+\gamma _0)  ( \gamma_5+ \gamma _3)& -1 & -1/2 \ \ 0 \ \  0
\end{eqnarray}
\caption{Parity   $P=1$ massive bosons.\label{massivesolPmend}}
\end{table}
%FIND OUT THE DIFFERENCES WITH THE MASSLESS CASE (?)
where the $\tilde 0$ subscript labels the solutions with negative  eigenvalue of $\frac{i}{2}\gamma_1\gamma_2$.
The solutions are classified with the aid of the normalized mass operator $H/M=\gamma_0$  and the
helicity $\frac{i}{2}\gamma_1 \gamma_2$ (this operator is obtained from the limiting case
$|{\bf k}|\rightarrow 0$ in eq. \ref{projeW}).
We note the solutions are mixed components of vector, antisymmetric and scalar components.
We  can also construct combinations with definite  properties under
the discrete transformations. Thus we find vectors  $\gamma_0\gamma_\mu$, axial vectors
 $\gamma_5\gamma_0\gamma_\mu$, scalars $\gamma_0$, and pseudo-scalars $\gamma_5\gamma_0$.
We obtain that the vectors become massive and  its longitudinal component becomes
physical. Just as in the massless case,
 an orthogonal  polarization basis can be defined.  For the pseudovector, its transverse and
longitudinal components are not physical.

On the other hand, the condition that they all belong to a vector representation forces us
to assume the antisymmetric and scalar terms  are in fact derivatives as in the massless case.
%Only the time derivative is different from zero when applied to
%solutions \ref{massivesol}
% One can simply reexpress these solutions in terms
%of the vectors in eq. \ref{ispossible}.
We note also there remain two
vector components constructed without internal spin, that is, constructed
from derivatives of scalar particles.  This structure is reminiscent  of  the
Higgs mechanism, in which massive vector fields absorb scalar degrees of
freedom due to breaking of the symmetry.

\section{Massless spin 1/2 particles }

We now show it is possible to give a Lorentz transformation  which describes fermions too.
This is done more naturally  in the context of a matrix equation  of the type of eq. \ref{Jaimeq},
and
whose solutions are bosons and fermions. This constitutes
progress
 in the  task of giving a unified description of these fields.
Indeed, we obtain
solutions that under Lorentz transformations of the form \ref{transfo}
one of the sides transforms trivially,   and therefore, we get spin $1/2$ objects transforming
 as the $(1/2,0)$ or $(0,1/2)$
 representations  of the  Lorentz group.

 The equation
 \begin{eqnarray}
\label {Jaimeferq}
  (1-\gamma_5)i \gamma_0\partial_\mu\gamma^\mu  {  \Psi}=0
\end{eqnarray}
has  this type of solutions.
The invariance algebra  of this equation contains the Lorentz generator
\begin{eqnarray}
\label{Lorentzchi}
J_{\mu\nu}^-=\frac{1}{2}(1-\gamma_5)J_{\mu\nu}=\frac{1}{2}(1-\gamma_5)[i(x_\mu\partial_\nu-x_\nu\partial_\mu)+
\frac{1}{2}\sigma_{\mu\nu} ]
\end{eqnarray}
(and  the other Poincar\'e generators).
Among the solutions of eq. \ref{Jaimeferq} we have again the $V$-$A$ vectors $u_i$
on Table   \ref{VmA}
which under the effect of $\frac{1}{2}(1-\gamma_5){\bf \Sigma\cdot{\hat{\bf  k} }}$
and $H/k_0=\frac{1}{2}(1-\gamma_5)\gamma_0\gamma^3$ lead to the
same quantum numbers.

The  Dirac operator in   eq. \ref{Jaimeferq} is defined on
a $2\times 2$ matrix space; nevertheless, the solutions lie in  the larger $4\times 4$ matrix space.
It is precisely this structure which leads to a set of solutions classified as
spin $1/2$ particles under  $J_{\mu\nu}^-$. Actually,
we have  as  additional symmetry of eq. \ref{Jaimeferq}  the group of linear complex transformations $G(2,C)$
with eight components, generated by
\begin{eqnarray}
\label{flavor}
\frac{1}{2} (1+ \gamma_5),\  \   f_{\mu\nu}=-\frac{i}{2}   (1+\gamma_5) \sigma_{\mu\nu},\end{eqnarray}
where   $\sigma_{\mu\nu}$ is given in eq. \ref{sigma}.
This means eq. \ref{Jaimeferq}
has   a freedom  in the choice  of the Lorentz
transformation since,  e.g., both $J_{\mu\nu}^-$ and $J_{\mu\nu}$
 are possible ones.
The  $\bfw_{i}$ terms on Table  \ref{eigenantis} are also a set of solutions of eq. \ref{Jaimeferq}.
However, their interpretation changes to fermions when using $J^-_{\mu\nu}$
to classify them.
Clearly, the nature of the solutions  depends on
the Hamiltonian and the set of transformations that we choose to classify them. But once our
choice is made, there is no ambiguity.
% We shall the assume  maximal symmetry classification hencefore. WHAT?

The unitary subgroups $SU(2)\times U(1)$  of the $G(2,C)$  symmetry operators in eq. \ref{flavor}
imply we have two additional  quantum numbers we can assign to
the  solutions.
In consideration that  this symmetry does not act on  the vector solution part,
and taking account of the known quantum numbers  of fermions  in nature  we shall  associate these
 operators with flavor and lepton number
respectively.
The  $SU(2)$ set of operators in \ref{flavor} leads  to a flavor doublet.  The $U(1)$ is in this case
not independent from the chirality. Choosing among the generators of
 $SU(2)$  $f_{30}$ to classify the solutions of eq. \ref{Jaimeferq},
these   are given on Table \ref{fermionsend},
\begin{table}
\begin{eqnarray}
Left-handed\  spin\ 1/2 \  particles &  \frac{1}{2}(1-\gamma_5)\gamma_0\gamma^3 & \frac{i}{4} (1-\gamma_5)\gamma_1\gamma_2  \ \ [f_{30},\ ] \nonumber \\
\bfw_{-1/2}({ k})  = \frac{1}{4} (1-\gamma_5) (\gamma_0+\gamma_3 ) & 1 & -1/2\ \ 1/2 \nonumber\\
\bfw_{-1/2}  ({\tilde   k})= \frac{1}{4} (1-\gamma_5)  (\gamma_1+i\gamma_2 )& -1 & 1/2\ \ 1/2 \nonumber \\
\hat \bfw_{-1/2}   ({ k})= \frac{1}{4} (1-\gamma_5 ) (\gamma_1-i\gamma_2 )& 1 & -1/2\  -1/2\nonumber\\ \nonumber
\hat \bfw_{-1/2}({ \tilde   k})  =  \frac{1}{4} (1-\gamma_5) (\gamma_0-\gamma_3 ) & -1 & 1/2\ \ -1/2
\end{eqnarray} %MAKE THEM RIGHT-HANDED TO RELATE TO Y acting on e^-_L ?{mass
\caption{Massless  fermions. \label{fermionsend}}
\end{table}
where the hat is used to distinguish  the flavor and in this case the product and commutator of $H$
and  $\frac{1}{2}(1-\gamma_5){\bf \Sigma\cdot{\hat{\bf  k} }}$ give the same results.
Eq. \ref{coordepw}  can be used to obtain the full coordinate dependence.
As  in the Weyl equation we obtain  solutions of a defined chirality
or helicity.
We have also negative energy solutions of the form \ref {coordepwneg} whose hermitian conjugates
are interpreted as right-handed  antiparticles. The latter could have been obtained by
departing  from an    equation with $V$+$A$ solutions.
In order to have a Dirac fermion and a fermion mass we need to
have an equation mixing both chirality  solutions. These shall be obtained
in Sections VII and VIII.

\noindent {\bf Gauge invariance}

We  prove  eq. \ref{Jaimeferq} has a gauge symmetry in the limiting
case of $\alpha_+\rightarrow 0$ in
\begin{eqnarray}
\label {Jaimeferqgauge}
[ \alpha_+ \frac {1}{2}(1+\gamma_5)
+\alpha_- \frac {1}{2}(1-\gamma_5) ]i \gamma_0\partial_\mu\gamma^\mu {  \Psi}=0.
\end{eqnarray}
Using   eqs. \ref{commugaugetrans}, \ref{commugaugeoth}
and \ref{genegaugeg},   \ref{genegaugeg5}, and commutation
relations with  $i\gamma_5 \gamma_0\partial_\mu\gamma^\mu $ it can be
shown
\begin{eqnarray}
\label{ferqgauge}
[(1-\gamma_5) \not\! \partial, (1+\gamma_5) \not\! \partial b(x)]_- & = &
2[\not\! \partial, \not\! \partial b(x)]_- -
2\gamma_5[\not\! \partial,  \not\! \partial b(x) ]_+ \\ \label{ferqgaugew}
[(1-\gamma_5) \not\! \partial, (1+\gamma_5) b(x)  \not\! \partial ]_- & = &
2[\not\! \partial, b(x) \not\! \partial ]_- -
2\gamma_5[\not\! \partial, b(x)\not\! \partial ]_+  \\ \label{ferqgaugex}
[(1+\gamma_5) \not\! \partial, (1-\gamma_5) \not\! \partial b(x)]_- & = &
2[\not\! \partial, \not\! \partial b(x) ]_- +
2\gamma_5[\not\! \partial,  \not\! \partial b(x) ]_+  \\  \label{ferqgaugez}
[(1+\gamma_5) \not\! \partial, (1-\gamma_5) b(x) \not\! \partial ]_- & = &
2[\not\! \partial, b(x)  \not\! \partial ]_- +
2\gamma_5[\not\! \partial, b(x)\not\! \partial ]_+ .
\end{eqnarray}
 The application of
the symmetry  generator
\begin{eqnarray}
\label{Gbs}
\bar G^b=[\alpha_-(1+\gamma_5)+\alpha_+(1-\gamma_5)]G^b
 \end{eqnarray}
on  eq. \ref{Jaimeferqgauge}, where $G^a$ is
given in eq. \ref{genegaugeg}, produces the terms in  eqs.
\ref{ferqgauge}-\ref{ferqgaugez} and
cancels the anticommutator contributions. Therefore, we obtain
 $\bar G^b$ will  be a generator of a gauge symmetry of eq. \ref{Jaimeferqgauge}
 in the sense explained before. In fact, in the  limit  $\alpha_+\rightarrow 0$
the symmetry is satisfied to first order since all other terms in  $e^{i\bar G^b}$
cancel. Unlike the case of
 eq. \ref{Jaimeqmle},  we note
only one gauge symmetry is allowed.

\section{Currents and    vertex interaction  }

From Noether's theorem we have a conserved current for each continuous symmetry present in the system.
 We  can construct, using eqs. \ref{Jaimeq}, \ref{Jaimeqdagger},
bilinear  current vector operators $j_\mu$, based on eq.  \ref{field algebra},   and
current vector expectation values    $tr(j_\mu)$, based on  eq. \ref{pointproduct}, satisfying
\begin{eqnarray}
\label{currentacon}
\partial^\mu j_\mu=0,\   \partial^\mu  tr(j_\mu)=0.
\end{eqnarray}
The form of the $j_\mu$ is similar to the  currents in Dirac's equation,
 given that some symmetries are shared by
both Dirac equation and eqs. \ref{Jaimeq}, \ref{Jaimeqdagger}.
In the case of eqs. \ref{Jaimeq},  \ref{Jaimeqmle}, and  \ref{Jaimeferq}
we also have the global symmetry  $\Psi\rightarrow e^{ia}\Psi$,
where $a$ is a  real parameter.
 The corresponding  current operator  is
\begin{eqnarray}
\label{currenta}
 j_\mu ={ \Psi}^\dagger\gamma_0\gamma_\mu { \Psi}.
\end{eqnarray}
This symmetry implies conservation of number of  particles, with
 the zero component of the current being positive definite so that it can
be interpreted as a probability density. This component has  already been
 considered when setting a normalization
condition  in eq. \ref{normalization}.

The  symmetry  $\Psi\rightarrow e^{ib\gamma_5}\Psi$, with $b$ real,  is valid for the massless equations
\ref{Jaimeqmle} and  \ref{Jaimeferq} and leads to the chirality current
\begin{eqnarray}
\label{currentb}
{ j_\mu }^{5}= { \Psi}^\dagger\gamma_5\gamma_0\gamma_\mu { \Psi}.
\end{eqnarray}

Expressions can be obtained also for the currents corresponding to the
energy-momentum tensor and the generalized angular momentum  which are equal
to those obtained for the Dirac equations. It is these which underlie
the classification of the solutions with $H$ and ${\bf \Sigma\cdot{\hat{\bf  k} }}$.
This partly justifies as well the classification done with the commutators of these operators  and the
solutions, given that they are
 also eigenfunctions under them.

The current operator corresponding to the gauge  symmetry in eq. \ref{Jaimeferqgauge}
(which overlaps with the above currents ${ j_\mu }$, ${ j_\mu }^{5}$) is given by
\begin{eqnarray}
\label{current}
{ j_\mu }^{gau}=  { \Psi} ^{\prime\dagger} \frac{1}{2} (1-\gamma_5)\gamma_0\gamma_\mu { \Psi}
\end{eqnarray}
(we take here the bra ${ \Psi} ^{\prime\dagger} $  possibly distinct from the ket  ${ \Psi} $).
Comparison of the current ${ j_\mu }^{gau}$
 with the form of the field
\begin{eqnarray}
\label{Amusec}
 A_{\mu}^-   ({k})= \frac{i}{2\sqrt{2}}(1-\gamma_5)\gamma_0\gamma_\mu e^{-i    kx}
\end{eqnarray}
derived from the $\bfu_{i}$ terms on Table  \ref{VmA}
strongly suggests a connection to the  transition matrix element
 of the $A_{\mu}^-$ operator field between
the two massless fermion solutions  ${ \Psi}^{\prime}$ and ${ \Psi}$
% Indeed,
%generally defining the tensor
%\begin{eqnarray}
%\label{tensor}
% \epsilon^{\mu\nu}=g^{\mu\nu} \epsilon_\nu \ \ \ {\rm (no\ \  sum)}
%\end{eqnarray}
%with $\epsilon_\nu$ containing  the spatial and polarization depedence, we may write generally write
%\begin{eqnarray}
%\label{Amusecgen}
% A_{5\mu}^-   ({k})= \epsilon_ {\mu }^{\  \nu} \frac{1}{2} (1-\gamma_5)\gamma_0\gamma_\nu.
%\end{eqnarray}
\begin{eqnarray}
\label{Amusecgentil}
 \langle  { \Psi}^{\prime}   |A_{\mu}^-   ({k})| { \Psi}\rangle.
\end{eqnarray}
Indeed, in QFT   the minimal coupling ${\mathcal{L}}=g j_\mu^{gau}A^{-\mu}$ in the Lagrangian implies
a vertex interaction which can lead to the
 expectation value  of the form
\begin{eqnarray}
\label{vertex}
g[\frac{1}{2}(1-\gamma_5)\gamma_0\gamma_\mu]_{\alpha \beta}\rightarrow
g\langle u(p_f,s_f)|\frac{1}{2}(1-\gamma_5)\gamma_0\gamma_\mu |u(p_i,s_i)\rangle,
\end{eqnarray}
where $(p_{i,f},s_{i,f} )$ are the initial and final momenta and  spins  of the fermions,   $k_\mu$
is the momentum of the vector field, and $g$ is the coupling constant.
A consistent interpretation of eqs. \ref{current}-\ref{Amusecgentil} is possible along these lines
by understanding $\langle  { \Psi}^{\prime}   |A_{\mu}^-   ({k})| { \Psi}\rangle$ as an interaction
with its
  assignment  to the vertex in eq. \ref{vertex}
and the coupling  constant $g=\frac{1}{\sqrt{2}}$.

(A more formal argument should take account of the exponential
factor in eq. \ref{Amusec} which can be done in the context of
QFT; it would lead, together with the space-time dependence of the
fermion wave functions to Dirac's delta
$(2\pi)^4\delta^4(k_i-p_{f}+p_{i})$.  Also, the substitution
$\ref{vertex}$ is one of many ways to obtain a contribution in a
perturbation expansion in terms of diagrams. Although
$A_{\mu}^-({k})$ should be properly normalized as a field of
units of $[energy]$ it is enough for our argument to keep the
polarization normalized).

\section{Lorentz $(3,1)$ Structure and Scalars  from $6$-$d$ Clifford Algebra}

In previous sections  we have derived a description of fermions and bosons
through equations implied by the structure of the Clifford algebra
in  $d=4$. Although the structure obtained is too simple to describe thoroughly aspects
of the SM (for example, the model cannot
include massive fermions) we have useful results which we would like to keep  as the prediction of interactions
in the form of  vertices relating vectors and fermions,
coupling constants, and
in particular,
hints of a description of isospin on left-handed particles. These features  are expected to
remain in higher dimensions, where we find a more elaborate structure.

The simplest generalization of the above model is to consider the six-dimensional Clifford algebra,
(the $d=5$ lives also in a $4\times 4$  space).
This is composed of  64 $8\times 8$  matrices and it
can be obtained as a tensor product of the original $4\times 4 $ algebra and the $2\times 2$ matrices
generated by the unit matrix  $1_2$
and the three  Pauli matrices, $\sigma_1$, $\sigma_2$, $\sigma_3$.
 We will use a basis for the $8\times 8$ matrix space in which we can  identify the underlying $d=4$ components
 \begin{eqnarray}
\label {generali}
\gamma_0\rightarrow \gamma_0^\prime=1_2  \otimes {\gamma_0},\ \gamma_1\rightarrow \gamma_1^\prime= 1_2\otimes  {\gamma_1} ,\
\gamma_2\rightarrow \gamma_2^\prime= \sigma_1\otimes {\gamma_2},\   \gamma_3\rightarrow \gamma_3^\prime=1_2 \otimes  {\gamma_3}.
\end{eqnarray}Then
  \begin{eqnarray}
\label {scalars}
1_8,\  \sigma_1\otimes 1_4,\ \gamma_5^\prime  = \sigma_2\otimes \gamma_2,\ \gamma_6^\prime= \sigma_3\otimes \gamma_2
\end{eqnarray}
 are 4-$d$ scalars since they
commute with the spin operators
\begin{eqnarray}
\label{sigmaprime}
 \sigma_{\mu\nu}^\prime=\frac{i}{2}[\gamma_\mu^\prime,\gamma_\nu^\prime] ,\   \mu=0,...,3,\    \nu=0,...,3.
\end{eqnarray}
In fact, the  matrices
$\gamma_\mu^\prime$ defined in eqs. \ref{generali} and \ref{scalars}
form
the $6$-$d$ Clifford algebra
 \begin{eqnarray}
\label {Clifford}
  \{\gamma_\mu^\prime,\gamma_\nu^\prime\} =2 g_{\mu\nu}.
\end{eqnarray}

As all $\gamma_\mu$ are generalized to $8\times 8$ matrices through a tensor product
 $\gamma_\mu\rightarrow 1_2\otimes \gamma_\mu $, $ \mu=0,...,3,$
without danger of ambiguity we shall use  a notation
in which we  now  assume that $\gamma_\mu$  represent
$8\times 8$ matrices.
We also use the quaternion-like notation  for the   representation of $1_2$ and the Pauli matrices in the $8\times 8$ matrix space
 \begin{eqnarray}
\label {quater}
1_8=1_2\otimes 1_4,\ I=\sigma_1\otimes 1_4,\   J= \sigma_2\otimes 1_4,\   K= \sigma_3\otimes1_4.
\end{eqnarray}
The $4$-$d$ algebra will be written in terms of
\begin{eqnarray}
\label {sustitution}
 \gamma_\mu^\prime=\gamma_\mu\ \ \mu=0,1,3\ \ \gamma_2^\prime=I \gamma_2,
\end{eqnarray}
and the scalars in eq. \ref{scalars} (here in Hermitian form) in terms of
\begin{eqnarray}
\label {sustitutionsca}
1=1_8,\ I,\ i\gamma_5^\prime=i J \gamma_2,\ i\gamma_6^\prime=i K \gamma_2.
 \end{eqnarray}
Because  $I$, $J$, $K$ commute with  $\gamma_2$,
it is possible to omit the tensor product sign. A more explicit form
of these matrices can be found in the appendix.
Then, all 64 elements of the $8\times 8$ algebra are obtained by  multiplying the 16 elements of
 the 4-$d$ algebra  generated by  the terms in eqs. \ref{sustitution} and
\ref{sustitutionsca}, and they can be written with this notation.
Hence, it will
be possible to identify every element constructed in this way in terms of  the 4-$d$ Lorentz representation
it belongs to.

The preceding definitions will also be applied for
the assignment $\gamma_5\rightarrow 1_2\otimes {\gamma_5}\equiv \gamma_5 $. Then, besides the scalar
elements of eq. \ref{scalars} (or eq. \ref {sustitutionsca}),
 we have the scalars
\begin{eqnarray}
\label {scalarspseu}
\gamma_5,\ I\gamma_5,\  J {\gamma_2}\gamma_5,\   K {\gamma_2}\gamma_5.
\end{eqnarray}
From these, $I \gamma_5$ commutes with these  and the scalar elements in
eq. \ref{sustitutionsca}.
 Excluding it and  the identity, the remaining six elements
generate an $SO(4)$ algebra, or equivalently, an $SU(2)\times SU(2)$ algebra.
The latter's generators consist of the right-handed elements
\begin{eqnarray}
\label {SU2r}
\frac{1}{4}(1+I \gamma_5)I,\ \frac{i}{4}(1+I \gamma_5)J {\gamma_2},\ \frac{i}{4}(1+I \gamma_5)K {\gamma_2},
\end{eqnarray} and left-handed elements
\begin{eqnarray}
\label {SU2}
I_1&=& \frac{i}{4}(1-I \gamma_5)J {\gamma_2}\\
 I_2&=&-\frac{i}{4}(1-I \gamma_5)K {\gamma_2}\\
I_3&=&-\frac{1}{4}(1-I \gamma_5)I \label{Iso3}.
\end{eqnarray}
The eight  form an $SU(2)_R\times SU(2)_L \times U(1)\times U(1)$, where the subscripts
$L$ and $R$  are added accordingly  (the normalization is chosen to
fit $\frac{1}{2}\sigma_i$).

\noindent{ \bf Chain breaking of $d=6$ Algebra}

The above symmetry operators immediately  show a close    connection to the
actual symmetries observed in nature at the massless level, that is,
 the $SU(2)_L$ of isospin and  $U(1)_Y$
of hypercharge groups. The eight scalars in eqs. \ref{sustitutionsca} and \ref{scalarspseu}, have a Cartan algebra
of dimension four, for which we can take the basis $1$, $I$, $\gamma_5$, $I\gamma_5$.    %CORREGIR
These operators can be arranged into the four projection operators
\begin{eqnarray}
\label {proje}
P_{++} &=&\frac{1}{4} (1+I \gamma_5)   (1+I)\\
P_{+-}&=&\frac{1}{4} (1+I \gamma_5)   (1-I)\\
P_{-+}&=&\frac{1}{4} (1-I \gamma_5)   (1+I)\\
P_{--}&=&\frac{1}{4} (1-I \gamma_5)   (1-I)
\end{eqnarray}
which, when combined with the Dirac operator, create the general massless Lorentz-invariant equation
\begin{eqnarray}
\label {geneq}
(\alpha_{++}P_{++} +\alpha_{+-}P_{+-} +\alpha_{-+}P_{-+} +\alpha_{--}P_{--} )
\gamma_0( i \partial^\mu{\gamma_\mu^\prime} ) \Psi=0 \ \mu=0,...,3.
\end{eqnarray}

We then have four different Lorentz-invariant degrees of freedom $\alpha_{++}$, $\alpha_{+-}$,
$\alpha_{-+}$, $\alpha_{--}$,
 for constructing a generalized
equation. One or various vanishing coefficients lead to degrees of freedom disappearing from the spectrum.
In fact,
the choice  of non-vanishing  coefficients  divides this equation into four classes.
For each class we assume that all fields
transform under the same Lorentz representation.
Additional conditions on the coefficients might lead to more symmetries to appear. The different choices
are as follows:

 In class I, only one coefficient is non-vanishing, e.g. %FIGURE
 $\alpha_{-+}\not=0$,  and $\alpha_{++}=\alpha_{+-}=\alpha_{--}=0$  (we will not consider the  different
 four permutations of the $\alpha_{ij}$ belonging
to this class and others, which have similar properties).
Without loss of generality here and in similar cases,
we may assume $\alpha_{-+}=1$.   %CHECK Here for the fact that the flavors are beyond $SU(2)$, etc
 This type of equation is similar to eq. \ref{Jaimeferq}, except that in this case, in addition to
the $U(1)$ gauge symmetry generated by $P_{-+}$,  we have a flavor $SU(6)$, whose elements
are projected by $P_{+-}+P_{-+}+P_{--}$.
%and the 4-$d$ Clifford algebra by the
%four $\gamma_\mu^\prime$. %QUOTE?
%(in all cases the right or left Casimir  is proportional to the $U(1)_L$ or
%$U(1)_R$ respectively)WHAT?

 In class II, in which
two $\alpha_{ij}$ vanish,  we have in general at least a $U(1)$ gauge symmetry and a
$SU(4)$ flavor symmetry. Furthermore, we consider three   possibilities for choices of the $\alpha_{ij}$.
In the  case $\alpha_{-+}=\alpha_{--}\not=0$ we have in particular
a $U(1)_L\times SU(2)_L$ gauge symmetry.
In this case, both fermions and vectors are obtained in the spectrum, but the fermions are all
left-handed  as for solutions  on Table \ref{fermionsend}, and their antiparticles right-handed.
The cases  $\alpha_{++}\not=0$, $\alpha_{-+}\not=0$, or $\alpha_{++}\not=0$, $\alpha_{--}\not=0$
 resemble eq. \ref{Jaimeferqgauge}
and break any possible gauge $SU(2)$ symmetry.

For class III, where only one $\alpha_{ij}=0$,  we have in general  an $SU(2)$ flavor symmetry, and
three gauge  $U(1)$s. In the case, $\alpha_{-+}=\alpha_{--}$ instead of one $U(1)$ we have also
 a gauge $SU(2)_L$ symmetry; by setting e.g.
$\alpha_{-+}=\alpha_{+-}$ or  $\alpha_{--}=\alpha_{+-}$ we get an equation which has a projection
of the form of eq. \ref{Jaimeqmle}, that is, with parity as a
 symmetry, a condition  necessary to have a solution of the form of an electromagnetic field.
The representations contain both vectors and fermions which are  both left-handed and right-handed.

 Finally, for class IV, in the case $\alpha_{++}=\alpha_{+-}=\alpha_{--}=\alpha_{-+}$    we have
a gauge $U(2)_L \times  U(2)_R$ and the representations only contain vectors of the type
appearing on Tables  \ref{VmA}, \ref{VpA} %CHECK
and \ref{eigenantis}, \ref{eigsend}. There is a possibility of finding a similar description to that of
class III if we use a Lorentz transformation projected with $L=P_{+-}+P_{-+}+P_{--}$. This case
will not be considered.

From the four choices described it is type III (or type IV under the condition stated) with
 $\alpha_{-+}=\alpha_{--}=\alpha_{+-}$  which can be parity conserving and which  contains
an $SU(2)_L$ symmetry. We shall associate this group with the  isospin and one  $U(1)$
with  the hypercharge. This case  is  analyzed in detail in the following section.

\section{Massless case: Type III spectrum, Unified $SU(2)\times U(1)$ }

We analyze  the equation
\begin{eqnarray}
\label {geneqIII} i L\gamma_0  \partial^\mu\gamma_\mu^\prime
\Psi=0,\ \mu=0,...,3,
\end{eqnarray}
where we use the projection operator
\begin{eqnarray}
\label {OpeYLor2N} L=P_{+-} +P_{-+} +P_{--}
=\frac{3}{4}-\frac{1}{4}(I+\gamma_5+I \gamma_5 ),
\end{eqnarray}
which corresponds to the type III case with
$\alpha_{+-}=\alpha_{-+}=\alpha_{--}$. The equation is invariant
under the set of Lorentz transformations
\begin{eqnarray}
\label {angularL}
J_{\mu\nu}^L=L[i(x_\mu\partial_\nu-x_\nu\partial_\mu)+\frac{1}{2}\sigma_{\mu\nu}^\prime]
,
\end{eqnarray}
where $\sigma_{\mu\nu}^\prime $ is defined in eq.
\ref{sigmaprime}. The scalar symmetries are classified into
flavor, with its generators  projected by $P_{++}$,
 an $SU(2)_L$   gauge symmetry, and  two other $U(1)$ gauge, according to
the arguments following  eq. \ref{Jaimeferqgauge}. We choose one
generator of $SU(2)_L$ to classify the solutions, say, $I_3$ in
eq. \ref{Iso3}, with eigenvalue $I_{s3}$. The other two $U(1)$
gauge generators, are chosen orthogonal to $I_3$ and will be taken
\begin{eqnarray}
\label {hyperch}
Y=-1+\frac{1}{2}(I+\gamma_5)% this minus is to make accord with eigenvalues... REMEMBER TO CHANGE Q SIGN
\end{eqnarray}
and $\tilde \gamma_5=L\gamma_5 $. $\tilde \gamma_5$ is
orthogonal to $Y$ and $I_3$ in the sense of $tr(\tilde \gamma_5
Y)=0$, and $tr(\tilde \gamma_5 I_3)=0$. The choice in eq.
\ref{hyperch} can be obtained from the demand  that the operator
lead to a gauge symmetry in the sense of eq. \ref{Gbs}. Another
justification  for these definitions will be given  later.
Although we  call it gauge we still need to prove that a chiral
symmetry as the $I_i$ actually is but we shall make this
assumption. There are also global symmetries which are related to
particle number conservation. $L$ is interpreted as the lepton
number, whose quantum number we denote by  $l$. The  Casimir of
$SU(2)_L$ $I_1^2+I_2^2+I_3^2$, with eigenvalue $I_s(I_s+1)$, is
not an independent component, but a linear combination of
$\tilde \gamma_5$, and $Y$.

We present the  fermion and boson solutions of eq.
\ref{geneqIII}  which we classify according to the  Hamiltonian and helicity projections, $L
\gamma_0\gamma^3$, $ \frac{i}{2} L I
\gamma_1\gamma_2$, the generator  $I_{3}$, and the quantum numbers, $Y$, $I_s$.
We also define the flavor as $f_{30}=\frac{1}{2}(1-L)\gamma_3
\gamma_0$ and its eigenvalue $f$.

{\bf Spin 1/2 particles}

 The $l=1$, $I_s=1/2$, $Y=-1$  massless fermions are given on Table \ref{doubletend},
\begin{table}
\begin{eqnarray} \nonumber
Left-handed\ massless\  spin\ 1/2\ particles,\ f=1/2 &L\gamma_0\gamma^3  & \frac{i}{2}  L I\gamma _1 \gamma _2\ \ \ \  I_{3} \\
\left( \begin{array}{c}
 \nu_{-1/2} (k) \\ \nonumber
l^-_{-1/2}(k)
 \end{array} \right )_L =
\left( \begin{array}{c}
 \frac{1}{8} (1-I\gamma _5  ) ( J \gamma_2-i K\gamma_2)   (\gamma_0+ \gamma_3) \\ \nonumber
\frac{1}{8} (1-I \gamma _5 )  ( 1+I)   (\gamma_0+ \gamma_3)
 \end{array} \right ) &
1 & -1/2
\begin{array}{c}  1/2 \\ -1/2 \end{array} \\
\left( \begin{array}{c}
\nu_{-1/2} (\tilde k)  \\
l^-_{-1/2} (\tilde k)
 \end{array} \right )_L =
\left( \begin{array}{c}
 \frac{1}{8} (1-I\gamma _5 ) ( J \gamma_2-i K\gamma_2) (\gamma_1+i I \gamma_2)  \\ \nonumber
\frac{1}{8} (1-I\gamma _5 ) ( 1+I) (\gamma_1+i I \gamma_2)
\end{array} \right ) & -1 & 1/2
\begin{array}{c} 1/2 \\ -1/2   \end{array} \\ \nonumber
 %%%%%%%%%%%%%%%%%%%%%%%%%"negative" energy terms %%%%%%%%%%%%%%%%%%%%%%%%%%
Left-handed\ massless\  spin\ 1/2\ particles,\ f=-1/2 &   L\gamma_0\gamma^3    & \frac{i}{2}  L I\gamma _1 \gamma _2\ \ \ \  I_{3} \\
\left( \begin{array}{c}
 \hat\nu_{-1/2} (k)   \\ \nonumber
\hat l^-_{-1/2}  (k)
 \end{array}\right )_L =
\left( \begin{array}{c}
 \frac{1}{8} (1-I\gamma _5 )  ( J \gamma_2-i K\gamma_2) (\gamma_1- iI \gamma_2) \\ \nonumber
\frac{1}{8} (1-I \gamma _5 )  ( 1+I)  (\gamma_1- iI \gamma_2 )
 \end{array} \right ) &
1 & -1/2
\begin{array}{c}  1/2 \\ -1/2 \end{array} \\ \nonumber
\left( \begin{array}{c}
 \hat{\nu}_{-1/2}(\tilde k)   \\ \nonumber
\hat l^-_{-1/2} (\tilde k)
 \end{array}\right )_L =
\left( \begin{array}{c} \frac{1}{8}  (1-I\gamma _5 ) ( J
\gamma_2-i K\gamma_2) (\gamma_0- \gamma_3)   \\ \nonumber
\frac{1}{8} (1-I \gamma _5 )  ( 1+I) (\gamma_0- \gamma_3)
 \end{array} \right ) &
-1 & 1/2
\begin{array}{c}  1/2 \\ -1/2  \end{array}
 \end{eqnarray}
 \caption{$l=1$, $I_s=1/2$, $Y=-1$  massless fermion multiplets in 5+1 $d$.\label{doubletend}}
\end{table}
where the  subscript  $- 1/2$ refers  to the value of the
helicity operator $[{\bf \Sigma\cdot\hat{\bf  p}},\ ]$, so that we also
present particles with opposite momentum, and the index $L$
denotes (in a redundant way) the left-handed character of the solutions.
The space-time dependence of these solutions can be obtained also
following eq. \ref{coordepw}. Negative
energy solutions are obtained by changing the sign of the
exponential  and the hermitian conjugates of the latter are
the antiparticle solutions.

 These spin 1/2 particles belong to the fundamental
representation of the non-abelian group $SU(2)_L$ and  are
labeled also by the $Y$ operator. In consideration of the
quantum numbers of leptons in nature, it follows we can associate
 $Y$ with the hypercharge and  the $I_i$ with the three generators
of isospin. We also associate the two elements distinguished only
by the $f$ quantum number (and a hat) with a  flavor doublet
which we identify with  any two lepton families among the three
generations, e. g., the left-handed electron and muon and their
neutrinos.

 Another part of the spectrum has positive chirality,
$l=1$, $I_s=0$, $Y=-2$ and  is given on Table \ref{singletend},
\begin{table}
\begin{eqnarray}\label{singlet}\nonumber \
Right-handed\ massless\  spin\ 1/2\ particles &
L\gamma_0\gamma^3  & \frac{i}{2}L I\gamma _1 \gamma _2\ \ [f_{30},\ ]
\\ \nonumber
 l^-_{1/2R}(k)= \frac{1}{8} (1+I \gamma _5 ) ( J \gamma_2-i K\gamma_2) \gamma_0 (\gamma_1+
i I \gamma_2)   & 1 &  1/2 \ \      1/2 \\ \nonumber l^-_{1/2R}
(\tilde k)=   \frac{1}{8} (1+I\gamma _5 ) ( J \gamma_2-i
K\gamma_2) \gamma_0 (\gamma_0+\gamma_3) & -1 & -1/2 \ \  1/2 \\
\nonumber
  \hat l^-_{1/2R}(  k) =\frac{1}{8}  (1+I\gamma _5 )  ( J \gamma_2-i K\gamma_2) \gamma_0
(\gamma_0-\gamma_3) & 1 &   1/2
    \ -1/2 \\ \nonumber
 \hat l^-_{1/2R} (\tilde k)= \frac{1}{8} (1+I\gamma _5 )  ( J \gamma_2-i K\gamma_2) \gamma_0 (\gamma_1-
i I\gamma_2) & -1 &  -1/2
 \ \   -1/2
\end{eqnarray}
\caption{$l=1$, $I_s=0$, $Y=-2$  massless fermion multiplets in
5+1 $d$.  \label{singletend}}
\end{table}
where    antiparticles can be obtained with the corresponding
transformations, and as in previous cases, the solutions
presented can be obtained from each other by a rotation.  The
quantum numbers correspond to  right-handed charged leptons, as
we will show, again in good correspondence with the SM.
%DEFINE l^-, etc.

\noindent{\bf Vectors}

The pure vector solutions are similar to the $\bfu_i$, $\bfut_i$
terms in  Tables  \ref{VmA}, \ref{VpA}. The isospin scalars can
be separated into their $V$+$A$ and $V$-$A$ components (all have
lepton number $l=0$, as required). The first are given on Table
\ref{righthandZend}.
\begin{table}[h]
\begin{eqnarray}
 \nonumber Vector \ solutions &  [H/k_0,\ ] &
[{\bf \Sigma\cdot\hat{\bf  p}},\ ] \\ \nonumber \tilde B_{1}(k)  =
\frac{1}{8} (1+I \gamma_5)(1-I )  \gamma_0(\gamma_1+i I \gamma_2)
& 2 & 1 \\ \nonumber \tilde B_{1}(\tilde k)=\frac{1}{8}  (1+I
\gamma_5)(1-I ) \gamma_0  (\gamma_1-i I \gamma_2)&  2 &  1 \\
\nonumber
\tilde B_{0}(k)=\frac{1}{8}   (1+I \gamma_5)(1-I )  \gamma_0(\gamma_0-\gamma_3) & 0 & 0 \\ \nonumber %THIS ONE HAS DIRAC EIGENVALUE 1
\tilde B_{0}(\tilde k) = \frac{1}{8}   (1+I \gamma_5)(1-I ) \gamma_0(\gamma_0+\gamma_3) & 0 & 0    %THIS ONE HAS DIRAC EIGENVALUE -1
\end{eqnarray}
\caption{$I_s=0$, $Y=0$, $V$+$A$  vectors  in 5+1 $d$.
\label{righthandZend}}.
\end{table}
 Comparing these solutions with those on Table  \ref{VpA}, we see they differ
by the substitution \ref{sustitution} and the projector $P_{+-}$.
Similarly, the $V$-$A$ terms can be obtained straightforwardly
from Table  \ref{VmA} and the projector $P_{-+}+ P_{--}$.  These
are given on Table \ref{BLend}.
\begin{table}
\begin{eqnarray}
 \nonumber Vector \ solutions &  [H/k_0,\ ]  &
[{\bf \Sigma\cdot\hat{\bf  p}},\ ] \\ \nonumber \label{BLbegin}
B_{-1}(k)= \frac{1}{4\sqrt{2} } (1-I \gamma_5) \gamma_0(\gamma_1-i I \gamma_2) & 2 & -1 \\ \nonumber
B_{-1}(\tilde k) =  \frac{1}{4\sqrt{2} } (1-I \gamma_5) \gamma_0(\gamma_1+i I \gamma_2) & 2 & -1 \\ \nonumber
B_{0}(k)=  \frac{1}{4\sqrt{2} } (1-I \gamma_5) \gamma_0(\gamma_0-  \gamma_3) & 0 & 0 \\ \nonumber
B_{0}(\tilde k) = \frac{1}{4\sqrt{2} } (1-I \gamma_5)
\gamma_0(\gamma_0+  \gamma_3)  & 0 & 0
\end{eqnarray}
\caption{$I_s=0$, $Y=0$, $V$-$A$  vectors in 5+1 $d$.
\label{BLend}}
\end{table}
 Taking account of the normalization, a combination of the terms $B_{i}$
 and $\tilde B_{i}$      can be taken which
carries the  hypercharge $Y$ in eq. \ref{hyperch}.
%SAY SOMETHING OF HYPECHARGE CURRENT AS
%the same vector space as %EXPLAIN MORE, or maybe before, when discussing currents
%the hypercharge current. ?  (question)
 We shall associate
this combination with the $B_\mu$ fields which carry the
hypercharge in
 the Weinberg-Salam model (Glashow, 1961, Salam, 1968, Weinberg, 1967).
%\cite{Weinberg}.

Three additional sets of solutions of the equation can be
described in terms of the fields $B_{i}$ on Table \ref{BLend} and
the generators of  $SU(2)_L$ in eqs. \ref{SU2}-\ref{Iso3}, which
are  written  in a spherical  basis on Table \ref{chm}.
\begin{table}[h]
\begin{eqnarray}
\nonumber
Isospin\ vector\ carriers & & I_{3}  \\ \nonumber
W^+_i(k) =&\frac{1}{\sqrt{2}}(J \gamma_2-i K \gamma_2)  B_{i}(k) &1 \label{chp} \\ \nonumber
W^0_i(k) =& I B_{i}(k)  &0 \\ \nonumber
W^-_i(k) =&\frac{1}{\sqrt{2}}( J\gamma_2+i K\gamma_2)B_{i}(k)
&-1
\end{eqnarray}
\caption{Isospin triplet   vector bosons  in 5+1 $d$.
\label{chm}}
\end{table}
%With this phase definition we have  $W^+_i=-{(W^-_i)}^\dagger$ (AS SUCH THIS IS NOT
%CORRECT. PUT ALSO INDEX? $\mu$?).
As these $V$-$A$ vector  solutions belong to the  adjoint
representation of group $SU(2)_L$, $I_s=1$, we associate them
with the fields $W_\mu^\pm$, $W_\mu^0$ of the electroweak theory.

\noindent{\bf Scalars and antisymmetric tensors}

The last part of the boson spectrum is composed of scalar and
antisymmetric $Y=-1$ doublets (and antiparticles). The solutions
are constructed similarly to the $\bfwt_i$  components on Table
\ref{eigsend} with  the addition of the factors containing
$I$, $ J \gamma_2$,  $ K \gamma_2$, which account for the
hypercharge  and isospin quantum numbers. The  corresponding
$Y=-1$  doublets  are given on Table \ref{nsend}.
\begin{table}
\begin{eqnarray}
\nonumber
Scalars\ and\ antisymmetric\ tensors\ & [H/k_0,\ ]   & [{\bf \Sigma\cdot}\hat{\bf  p},\ ]\ \ I_{3} \\
\left( \begin{array}{c}
 \tilde n_0(k)\\ \nonumber
\tilde v_0(k)
 \end{array}\right )=
\left( \begin{array}{c}
 \frac{1}{8} (1+I \gamma _5) (J \gamma_2- i K \gamma_2)  ( \gamma_0 +\gamma _3)\\
 \frac{1}{8} (1+I \gamma _5) ( 1-I )  ( \gamma_0 +\gamma _3) \end{array} \right ) &2&0
\ \begin{array}{c} 1/2    \\ \nonumber -1/2  \end{array}
 \\
\left( \begin{array}{c}
 \tilde n_{0}(\tilde k)  \\ \nonumber
\tilde v_{0}(\tilde k)
 \end{array}\right )=
\left( \begin{array}{c}
 \frac{1}{8}  (1+I \gamma _5) (J \gamma_2- i K \gamma_2)  ( \gamma_0 -\gamma _3)\\ \nonumber
 \frac{1}{8} (1+I \gamma _5) ( 1-I )  ( \gamma_0 -\gamma _3)
 \end{array}\right ) &2&0 \ \begin{array}{c} 1/2    \\ -1/2  \end{array}     \\ \nonumber
\left( \begin{array}{c}
 \tilde n_{ 1}(k)  \\ \nonumber
\tilde v_{ 1}(k)
 \end{array}\right )=
\left( \begin{array}{c}
  \frac{1}{8} (1+I \gamma _5) ( J \gamma_2- i K\gamma_2)  ( \gamma_1 +i I\gamma _2)\\ \nonumber
 \frac{1}{8} (1+I \gamma _5)( 1-I )  ( \gamma_1 +iI\gamma _2)
 \end{array}\right )&0& 1  \ \begin{array}{c} 1/2    \\ -1/2  \end{array}      \\ \nonumber
\left( \begin{array}{c} \tilde  n_{ 1}(\tilde k)  \\ \nonumber
\tilde v_{ 1}(\tilde k)
 \end{array}\right )=
\left( \begin{array}{c}
 \frac{1}{8} (1+I \gamma _5)  ( J \gamma_2- i K \gamma_2)  ( \gamma_1 -i I \gamma _2)\\ \nonumber
 \frac{1}{8} (1+I \gamma _5)  ( 1-I )  ( \gamma_1 -i I \gamma _2)\end{array}\right )&0& 1
 \ \begin{array}{c} 1/2    \\ -1/2  \end{array}
\end{eqnarray}
\caption{$I_s=1/2$, $Y=-1$, boson chiral terms  in 5+1 $d$.
\label{nsend}}
\end{table}
The same problems arise regarding   the  Lorentz interpretation
of antisymmetric terms as for  Tables \ref{eigenantis},
\ref{eigsend}. The same procedure in extracting from these
solutions  vector and scalar components can be used.  Again here
there is a parallelism with the SM. A scalar particle appears in
a doublet and we will see it is involved in giving  masses to the
particles. For this reason, we may associate this degree of
freedom with the Higgs particle. We   leave open the question of
whether these mass terms can be obtained from a gauge
transformation, although the form of the proposed gauge
transformation here suggests it should be possible.

Summarizing,  the positive energy solutions are   the vectors
$B_i$ and $\tilde B_i$ which amount to eight degrees of freedom,
where we are taking  account of both directions of momenta for
given helicity. The isospin vectors $W_i^{\pm,0}$  have twelve
degrees of freedom and the antisymmetric tensors and scalars
$\tilde n_i$ and $\tilde v_i$ eight, and with their antiparticles sixteen.
These add up to thirty-six bosons. We have obtained massless spin
1/2 particles in a doublet and a singlet; these use four and two
degrees of freedom respectively. Taking account of antiparticles
and the two flavors we have twenty-four fermion degrees of
freedom.  Altogether, these add up to sixty degrees of freedom.
The reason for not having sixty-four active is the four inert
degrees of freedom projected by $P_{++}$, which are not
influenced by  the Hamiltonian, projected by $L$.

 %COMMENT ON THE FACT THAT IT APPEARS AS IF THE SYMMETRY WAS BROKEN

\section{Massive case: Symmetry breaking of $SU(2)\times U(1)$ }

 In seeking a  massive extension of eq. \ref{geneqIII}
we expect all the hermitian combinations of the
 scalar terms  in eqs. \ref{sustitutionsca}
  and \ref{scalarspseu}, multiplied by $\gamma_0$ (in a Hamiltonian form of the equation),
 to be scalars with respect to the Lorentz transformation
 \begin{eqnarray}
\label{angularpri} J_{\mu\nu}^\prime=
i(x_\mu\partial_\nu-x_\nu\partial_\mu)+\frac{1}{2}\sigma_{\mu\nu}^\prime,
\end{eqnarray}
which just generalizes eq. \ref{angular}. However, if we also
demand that they be scalars  with respect to $ J_{\mu\nu}^L$ in
eq. \ref{angularL}, then the choices are reduced to
\begin{eqnarray}
\label{massterms}
M_1&=&\frac{M}{2}(1-I), \\
 M_2&=& \frac{i M}{2}(\gamma_5-I \gamma_5),\\
 M_3&=& -\frac{M}{2}J\gamma_2  (1+\gamma_5), \label{M3} \\
M_4&=& \frac{M}{2} K\gamma_2  (1+\gamma_5), \label{masstermsend}
\end{eqnarray}
where $M$ is the mass constant. Now, the only non-trivial scalar
that commutes with   all  $M_i$ terms is $L$. Nevertheless, if we
relax this condition we obtain in addition that precisely and only
\begin{eqnarray}
\label{charge} Q=I_3+\frac{1}{2} Y
\end{eqnarray}
commutes with $M_3$ and $M_4$ ($Q^\prime=I_3-\frac{1}{2} Y$
commutes with $M_1$ and $M_2$). As $Q$ is the electric charge we
deduce the electromagnetic $U(1)_{em}$ remains a symmetry while
the hypercharge and isospin  are broken. We stress that   $Q$ is
deduced, rather than being imposed, as the only  additional
symmetry consistent with massive terms. $M_3$ and $M_4$ do not
commute among themselves, but rather, can be obtained from each
other through a  unitary transformation involving $\gamma_5$. We
therefore choose one, $M_3$,  to study the massive
representations. We will show the  equation
%PUT NOTATION WHERE ALSO   $\gamma_0$ APPEARS in M_4 ?
\begin{eqnarray}
\label {eqIIImass} (L\gamma_0 i \partial^\mu\gamma_\mu^\prime  -
M_3\gamma_0){  \Psi}=0\ \  \mu=0,...,3,
\end{eqnarray}
%check that all "\gamma_5 I" I \gamma_5 have the same order
gives rise to massive and massless fermions and vectors that are
contained  in the SM, at symmetry breaking.

\noindent{\bf Vectors}

Despite the presence of a massive term  we get a set of vector
components which remain massless, as their product with the mass
term $M_3$ (or $M_4$)  in eq. \ref{eqIIImass} vanishes. These are
the combination of the  massless and chargeless terms on Tables
\ref{BLend} and \ref{chm}
\begin{eqnarray}
\label {photon}
A_{Li}=\frac{1}{\sqrt{2}}(B_{i}-W_i^0). %CHECK NORMALIZATION
\end{eqnarray}

There are several parity operators for eq. \ref{eqIIImass} with
the necessary properties.  They differ by the square which leads
to different projection operator combinations. The only one
leading to non-trivial solutions  acts on the  same space as  $Q$
and is of the same rank.  This is
\begin{eqnarray}
\label{parity} P=M_3\gamma_0 \wp,
\end{eqnarray}
where $\wp$ is defined as for eq. \ref{P} and $M_3$ is given in
eq. \ref{M3}.

The remaining bosons  become massive. The massive chargeless
solutions are a combination of the
 vectors $B_{ i}$, $\tilde B_{ i}$  on Tables \ref{righthandZend}, \ref{BLend}, the $W_{i}^0$ on Table
\ref{chm}, the $\tilde n_i$ bosons on  Table \ref{nsend}, and
their antiparticles $n_i$, with  $n_{0}(k)=\tilde n_{0}^\dagger(\tilde k)$, $n_{0}(\tilde k)=\tilde n_{0}^\dagger(k)$,
$n_{-1}( k)=-\tilde n_{1}^\dagger( k)$, $n_{-1}(\tilde k)=-\tilde n_{1}^\dagger(\tilde k)$. We construct the latter using Table
\ref{eigenantis} and multiplying on the left by operators
carrying the isospin and hypercharge, with phases as on Table
\ref{nsend}. The solutions   can be classified in  two groups,
depending on the value of the commutator $ [ M_3 \gamma_0,{
\Psi}]$, or equivalently, by the value of the operator $P$. When
the commutator is zero we have the solutions on Table
\ref{vectormassend}.
\begin{table}
\begin{eqnarray}
\label{vectormass}\nonumber
Massive\  bosons &  M_3 \gamma_0 /M& \frac{i}{2}LI\gamma_1\gamma_2 \ \ [ H/k_0,\ ]   \ \  [{\bf \Sigma\cdot{\hat{\bf  k} }} ,\ ] \\
{{ P}}_ {1} (M,{\bf 0}) = \frac{1}{2}( \tilde  B_{ 1}(k) +
 \frac{1}{\sqrt{2}}(B_{-1} (\tilde k)
+ W_{-1}^0(\tilde k) )\nonumber \\ \nonumber
-  \tilde n_1( k)+
n_{-1}(\tilde k) )
& 1 & 1/2 \ \ 0\ \ 1\\ \nonumber
{{ Q}}_ {1} (M,{\bf 0})= \frac{1}{2}(   \tilde B_{ 1}(  k) +
\frac{1}{\sqrt{2}}(B_{-1}(\tilde k) + W_{-1}^0(\tilde k) )
\nonumber \\ \nonumber +  \tilde n_1( k)-  n_{-1}(\tilde k) )
& -1 & 1/2\ \ 0 \ \ 1 \\ \nonumber
{{ P}}_ {-1} (M,{\bf 0})= \frac{1}{2}(   \tilde B_{ 1}(\tilde k)+
\frac{1}{\sqrt{2}}(B_{-1}(  k) + W_{- 1}^0(  k) ) \nonumber \\ \nonumber -
\tilde n_1(\tilde k)+  n_{-1}(k) )
& 1 & -1/2\ \ 0\ \ -1 \\ \nonumber
{{ Q}}_ {-1} (M,{\bf 0})= \frac{1}{2}(   \tilde B_{ 1}(\tilde k)+
\frac{1}{\sqrt{2}}(B_{-1}(  k) + W_{- 1}^0(  k) ) \nonumber \\ \nonumber +
\tilde n_1(\tilde k)-  n_{-1}(  k) )
& -1 & -1/2\ \ 0 \ \ -1 \\ \nonumber
%%%%%%%%%%%%%%%%%%%%%%%%%%%%here start longitudinal%%%%%%%%%%%%%%%%%%%%%%%%%%%%
{{ P}}_ {0} (M,{\bf 0}) = \frac{1}{2}(  \tilde  B_{0}(k)+
\frac{1}{\sqrt{2}}(B_{0}(\tilde k)
 +W_{0}^0 (\tilde k))-   \nonumber \\ \nonumber \tilde n_0(k)-  n_0(\tilde k)
) & 1 &  1/2 \ \ 0\ \  0
 \\ \nonumber
{{ Q}}_ {0} (M,{\bf 0})= \frac{1}{2}(  \tilde  B_{0}(k)+
\frac{1}{\sqrt{2}}(B_{0} (\tilde k) + W_{0}^0 (\tilde k))+
\nonumber \\ \nonumber  \tilde n_0(k)+  n_0(\tilde k) ) & -1 & 1/2 \ \ 0\ \  0
 \\ \nonumber
{{ P}}_ {\tilde 0} (M,{\bf 0}) = \frac{1}{2}(  \tilde
B_{0}(\tilde k)+ \frac{1}{\sqrt{2}}(B_{0} (   k) + W_{ 0}^0
(   k) )- \nonumber \\ \nonumber \tilde n_0(\tilde k)-  n_0(k) ) & 1 &
-1/2 \ \ 0\ \  0
 \\ \nonumber
{{ Q}}_ {\tilde 0} (M,{\bf 0})= \frac{1}{2}(  \tilde
B_{0}(\tilde k)+ \frac{1}{\sqrt{2}}(B_{0}(  k) + W_{ 0}^0
(   k) )+ \nonumber \\  \nonumber \tilde n_0(\tilde k)+ n_0(k) ) &
-1 & -1/2 \ \ 0\  \ 0
\end{eqnarray}
\caption{$P=1$ massive bosons. \label{vectormassend}}
\end{table}
%NEED TO add third quanTUM NUMBER TO DISTINGUISH LAST FOUR
The $k$ and $\tilde k$ arguments simply label the vector
components (in the massless solutions) in terms of which the
massive solutions are constructed. The non-zero terms for the
commutator with $M_3 \gamma_0$ are given on Table \ref{vectormassMasse}.
\begin{table}
\begin{eqnarray}
\nonumber Massive\ bosons  &  M_3
\gamma_0/M & \frac{i}{2}LI\gamma_1\gamma_2 \ \ [ H/k_0,\ ]   \ \
  [{\bf \Sigma\cdot{\hat{\bf  k} }} ,\ ] \\
{\bar { P}}_ {1} (M,{\bf 0}) = \frac{1}{2}( \tilde  B_{ 1}(k) -
\frac{1}{\sqrt{2}}(B_{-1} (\tilde k) + W_{-1}^0(\tilde k) )+
\nonumber \\   \nonumber  \tilde n_1( k)+  n_{-1}(\tilde k) )
& 1 & 1/2 \ \ 2\ \ 1\\
{\bar { Q}}_ {1} (M,{\bf 0}) = \frac{1}{2}(   \tilde B_{ 1}(  k)
- \frac{1}{\sqrt{2}}(B_{-1}(\tilde k) + W_{-1}^0(\tilde k) )-
\nonumber \\   \nonumber \tilde n_1( k)-  n_{-1}(\tilde k) )
& -1 & 1/2\ \ -2\ \ 1 \\
{\bar { P}}_ {-1} (M,{\bf 0}) = \frac{1}{2}(   \tilde B_{
1}(\tilde k)- \frac{1}{\sqrt{2}}(B_{-1}(  k) + W_{- 1}^0(  k) )+
\nonumber \\   \nonumber \tilde n_1(\tilde k)+  n_{-1}(k) )
& 1 & -1/2\ \ 2\ \ -1 \\
{\bar { Q}}_ {-1} (M,{\bf 0}) = \frac{1}{2}(   \tilde B_{
1}(\tilde k)- \frac{1}{\sqrt{2}}(B_{-1}(  k) + W_{- 1}^0(  k) )-
\nonumber \\  \nonumber  \tilde n_1(\tilde k)-  n_{-1}(  k) )
& -1 & -1/2\ \ -2\ \ -1 \\
%%%%%%%%%%%%%%%%%%%%%%%%%%%%here start longitudinal%%%%%%%%%%%%%%%%%%%%%%%%%%%%
{\bar { P}}_ {0} (M,{\bf 0}) = \frac{1}{2}(  \tilde  B_{0}(k)-
\frac{1}{\sqrt{2}}(B_{0}(\tilde k)
 +W_{0}^0 (\tilde k))+ \nonumber \\  \nonumber   \tilde n_0(k)-  n_0( \tilde k)
) &  1 &  1/2 \ \ 2\ \  0
 \\
{\bar { Q}}_ {0} (M,{\bf 0}) = \frac{1}{2}(  \tilde  B_{0}(k)-
\frac{1}{\sqrt{2}}(B_{0} (\tilde k) + W_{0}^0 (\tilde k))-
\nonumber \\  \nonumber \tilde n_0(k)+  n_0(\tilde k) ) & -1 & 1/2 \ \ -2\ \  0
 \\
{\bar { P}}_ {\tilde 0} (M,{\bf 0}) = \frac{1}{2}(  \tilde
B_{0}(\tilde k)- \frac{1}{\sqrt{2}}(B_{0} (   k) + W_{ 0}^0
(   k) )+ \nonumber \\ \nonumber  \tilde n_0(\tilde k)-  n_0(k) ) &  1 &
-1/2 \ \ 2\ \  0
 \\
{\bar { Q}}_ {\tilde 0} (M,{\bf 0}) = \frac{1}{2}(  \tilde
B_{0}(\tilde k)- \frac{1}{\sqrt{2}}(B_{0}(  k) + W_{ 0}^0
(   k) )-\nonumber \\  \nonumber \tilde n_0(\tilde k)+ n_0( k) )
& -1 & -1/2 \ \ -2\  \ 0  %eryMa4RenP[15]
\end{eqnarray}
%Explain more
\caption{$P=-1$ massive bosons. \label{vectormassMasse}}
\end{table}

We have also a set of charged vector particles, constructed from
the $W_\mu^\pm$ on Table  \ref{chm}  and the charged
doublet components $\tilde v_i$ on Table \ref{nsend}, and their
antiparticles.
  The $Q=-1$ components are given on Table \ref{WsMassend},
\begin{table}[h]
\begin{eqnarray}
\label{WsMass}\nonumber Massive\ \ \ W's  &  M_3 \gamma_0/M &
\frac{i}{2}LI\gamma_1\gamma_2 \ \ [ H/k_0,\ ]   \ \
  [{\bf \Sigma\cdot{\hat{\bf  k} }} ,\ ] \\ \nonumber
 W_{M1}^-=\frac{1}{\sqrt{2}}( W^-_{-1}(\tilde k)-\tilde v_{1 }(k)  )
& 1 & 1/2 \ \ 1\ \  1 \\ \nonumber %eryMa4RenP[9]
\hat W_{M1}^-=\frac{1}{\sqrt{2}}( W^-_{- 1}(\tilde k) +\tilde
v_{1 }(k)  )
& -1 & 1/2 \ \   -1 \ \ 1  \\ \nonumber %eryMa4RenP[11]
W_{M-1}^-=\frac{1}{\sqrt{2}}( W^-_{-1} (k) -\tilde v_{1 } (\tilde
k)  )
   & 1 & -1/2 \ \  1\ \ -1 \\ \nonumber %eryMa4RenP[13]
\hat W_{M-1}^-=\frac{1}{\sqrt{2}}( W^-_{-1 }(k) +\tilde v_{  1 }
(\tilde k)  )
  & -1 & -1/2 \ \ -1\ \  -1  %eryMa4RenP[15]
\\ \nonumber
%%%%%%%%%%%%%%%%%%%%%negative mass %%%%%%%%%%%%%%%%%%%%%%%
 W_{M0}^-=\frac{1}{\sqrt{2}}(W^-_0(\tilde k) -\tilde v_{0 } ( k) )
&  1 & 1/2 \ \ 1\ \  0 \\ \nonumber %eryMa4RenP[9]
\hat W_{M0}^-=\frac{1}{\sqrt{2}}( W^-_{0}(\tilde  k) +\tilde v_{  0
}(k)  )
& -1 &  1/2 \ \   -1 \ \ 0  \\ \nonumber %eryMa4RenP[11]
W_{M\tilde 0}^-=\frac{1}{\sqrt{2}}(W^-_{0}( k) +\tilde v_{0
}(\tilde k)  )
   &  1 & -1/2 \ \  1\ \ 0 \\ \nonumber %eryMa4RenP[13]
\hat W_{M \tilde 0}^-=\frac{1}{\sqrt{2}}( W^-_{0 } ( k)
-\tilde v_{0 }(\tilde k)  )
  & -1 & -1/2 \ \ -1\ \ 0     %eryMa4RenP[15]
\end{eqnarray}
\caption{Charged $Q=-1$  massive vector bosons. \label{WsMassend}}
\end{table}
where the $\tilde 0$ subscript labels the solution with negative
eigenvalue of $\frac{i}{2}LI\gamma_1\gamma_2$. The positively
charged terms can be obtained from $(W^-_{Mi})^\dagger$.
%We note the $W$ becomes a particle acting on both chirality  %BULLSHIT
%components $(1+\gamma_5)/2$ and $(1-\gamma_5)/2$, unlike the massless
%case where   $W$ acts only on negative chirality terms. %CHECK THIS STATEMENT

\noindent {\bf Spin 1/2 particles}

The application of the massive terms $M_3$ and  $M_4$ in eqs.
\ref{M3} and \ref{masstermsend} to  the $Y=-1$ $I_{s3}=1/2,$ (with $Q=0$),
``neutrino" elements  and their antiparticles gives zero, which
implies
 the neutrinos remain massless. In addition, %CHECK IF USE OF WORD IS OK
the neutrino  and antineutrino solutions lack  a right-handed and
left-handed partner respectively to be able to form   Dirac
massive particles.

On the other hand, the massive term  $M_3$  (or  $M_4$) breaks
the chiral symmetry  mixing values of chirality and causing
 the charged fermions to acquire a mass.   The lepton number
$l=1$,  and charge $Q=-1$ wave functions are given on Table \ref{eleMassend}.
\begin{table}[h]
\begin{eqnarray}
\nonumber
Charged\  massive\  spin\ 1/2\  particles\ &  [M_3 \gamma_0/M, ] & [\frac{i}{2} LI  \gamma_1\gamma_2,\ ]   \ \  [f_{30},\ ]    \\ \nonumber
 u^-_{1/2} =\frac{1}{\sqrt{ 2} }(l^-_{-1/2 L}(\tilde k) -l^-_{1/2 R} (k))
& 1 & 1/2  \ \   1/2  \\ \nonumber
 v^-_{1/2} =\frac{1}{\sqrt{  2 }}(l^-_{-1/2 L} (\tilde k) +l^-_{1/2 R} (k) )
& -1 & 1/2  \ \    1/2  \\ \nonumber
u^-_ {-1/2} =\frac{1}{\sqrt{  2 }}(l^-_{-1/2 L}(k)  -l^-_{ 1/2 R}
(\tilde k) )
& 1 & -1/2 \ \   1/2  \\ \nonumber
 v^-_{-1/2} =\frac{1}{ \sqrt{ 2 }}(l^-_{-1/2 L}(k)  +l^-_{ 1/2 R} (\tilde k) )
& -1 & -1/2 \ \   1/2  \\ \nonumber
\hat u^-_{1/2} =\frac{1}{ \sqrt{ 2 }}(\hat l^-_{-1/2 L} (\tilde
k) -\hat l^-_{1/2 R}(k) )
& 1 & 1/2 \ \   -1/2  \\ \nonumber
\hat v^-_{1/2} =\frac{1}{\sqrt{  2 }}(\hat l^-_{-1/2 L} (\tilde
k) +\hat l^-_{1/2 R}(k) )
& -1 & 1/2 \ \   -1/2  \\ \nonumber
\hat u^-_ {-1/2} =\frac{1}{\sqrt{  2} }(\hat l^-_{-1/2 L}(k)
-\hat l^-_{1/2 R} (\tilde k) )
& 1 & -1/2 \ \   -1/2  \\ \nonumber
\hat v^-_{-1/2} =\frac{1}{\sqrt{  2} }(\hat l^-_{-1/2 L}(k)
+\hat l^-_{1/2 R} (\tilde k) ) & -1 & -1/2 \ \   -1/2
\end{eqnarray}
\caption{Charged  $Q=-1$ massive fermions. \label{eleMassend}}
\end{table}
The charge of these fermions  leads to their association  with
any two of   the negatively   charged massive leptons $e^-$,
$\mu^-$, or $\tau^-$.

%{\bf Vertices}

\section{Relation to physical fields}

There remains to classify  the vector fields obtained  in  the
breaking of the $SU(2)_L\times U(1)_Y$ to the $Q$ symmetry,
according to the discrete symmetries.
 The terms found, $A_{Li}$
in eq. \ref{photon},  and $P_i$, $\bar P_i$, $Q_i$, $\bar Q_i$ on
Tables \ref{vectormassend}, \ref{vectormassMasse}, sum twenty
degrees of freedom. Similar combinations as for the massive
vector terms $U_{i}$, $\bar U_{i}$, $V_{i}$, $\bar V_{i}$ on
Tables \ref{massivesol}, \ref{massivesolPmend} can be taken to
obtain terms with the necessary transformation properties.

%The most straightforward assignment we can make is to relate
%the  massless chargeless vector $A_{Li}$  in  eq. \ref{photon} to the photon.
%However, these terms describe only one type of chirality. The
%other chirality is contained in the massive terms in eq. \ref{vectormass}.
%As we the $U(1)_Q$ gauge symmetry remains valid
%even in the massive case, we expect these degrees of freedom to
%remain massless, although it appears this can only be achieved by taking
%a complete series of diagrams.

%With the caveat that these terms do not
%necessarily represent solutions to eq. \ref{eqIIImass},
%the vector components of the photon are obtained by extending
%the solutions  $A_{Li}$=\, to an  odd parity
%vector field.

%CORRECT THIS

%Without expecting that solutions be eigenvalues of the mass term we construct a
%massless solution  by considering only the kinetic term in eq. \ref{geneqIII}. This equation
%is parity invariant, a symmetry satisfied by the electromagnetic field.

The (normalized) vector components solutions of Tables
\ref{vectormassend}, \ref{vectormassMasse}, which transform as a
non-axial vector by $P$ in eq. \ref{parity} are given by
\begin{eqnarray}
\label{photoncompl} A_\mu=\frac{1}{2}Q\gamma_0\gamma_\mu.
\end{eqnarray}
$A_\mu$  can be represented as a mixture of two chargeless  and
massless  components.
 On  the one hand,   Tables \ref{vectormassend}, \ref{vectormassMasse} contain the  special combination, of the $B_{i}$
on Table \ref{righthandZend},  and the $\tilde B_{i}$ on Table
\ref{BLend}, which form precisely
\begin{eqnarray}
\label{Zcompl} B_\mu=\frac{1}{2\sqrt{3}}Y \gamma_0\gamma_\mu,
\end{eqnarray}
 that is, the hypercharge carriers.
 This gives   another  justification for the choice of $Y$ given in eq. \ref {hyperch},   which is the
operator giving the correct values for the hypercharge of
fermions. Thus we obtain another  argument  needed to set $Y$
whose background is  in   the way we arrive at the expression for
$Q$ in eq. \ref{charge}.
%We note that in this field the twoto
%different quantum numbers which characterize the interaction with
% the different chiral fermion components
%imply different occupation probabilities for the two field helicities. %CHECK WHETHER
% A  VECTOR HELICITY IS REALLY RELATED TO FERMION HELICITY.
On the other hand, we can extract the chargeless vector
components for the isospin triplet from Table \ref{chm}
\begin{eqnarray}
\label{Wcompl}
W_\mu^0=  I_3\gamma_0\gamma_\mu, %CHECK NOTATION
\end{eqnarray}
where $I_3$ is given in eq. \ref{Iso3}.

From the expression for $Q$ and eqs. \ref{photoncompl},
\ref{Zcompl}, and  \ref{Wcompl} we easily obtain
\begin{eqnarray}
\label{Weinangle}
A_\mu=\frac{1}{2}W_\mu^0+\frac{\sqrt{3}}{2}B_\mu.
\end{eqnarray}
The value of Weinberg's angle $\theta_W$  is derived  immediately
from this equation by making an analogy with the new fields
obtained in the  SM, after application of the Higgs mechanism.
The photon then has the form
\begin{eqnarray}
\label{photonSM} A_\mu=\frac{1}{\sqrt{
{g^\prime}^2+g^2}}(gB_\mu+g^\prime W^0_\mu),\end{eqnarray} where
$g$ and $g^\prime$ are respectively the isospin and hypercharge
coupling constants. We obtain  $\frac{g^\prime}{g}
=\frac{1}{\sqrt{3}}$. As in the SM
$tan(\theta_W)=\frac{g^\prime}{g}$ we find
 \begin{eqnarray}
\label{sinWeinangle} sin^2(\theta_W)=.25
\end{eqnarray}

The $Z_\mu$ field can be constructed by considering the
orthogonal combination to $A_\mu$ in eq. \ref{Weinangle}
\begin{eqnarray}
\label{MassZ} Z_\mu=\frac{\sqrt{3}}{2}W_\mu^0-\frac{1}{2}B_\mu.
\end{eqnarray}
 We therefore find the $A_\mu$ and $Z_\mu$ span
the vector components on Tables
\ref{vectormassend}, \ref{vectormassMasse}.

The charged massive solutions can be related to the $W^\pm_\mu$
components in the SM. Although we obtained a difference in the
masses  of the $Z_\mu$ and $W_\mu^\pm$ this is not the
corresponding to the one obtained in the SM. Also, the vector
particle $A_\mu$ is massive.  We attribute these differences to
the fact that term  $M_3\gamma_0$  does not commute with the
kinetic term in eq. \ref{geneqIII} which is required to allow for
simultaneously massive and massless solutions in the space
projected by $Q$. Indeed, the space spanned by the vectors
$A_{Li}$ in eq. \ref{photon} is annihilated by $M_3$. This fact
allows them to be  massless solutions.

\noindent{\bf Coupling constants: Vector fermion-current vertices}

 Following the
steps which allow for  a  vertex interpretation of   eq. \ref
{Amusecgentil}
 it is possible to derive the
vertices  describing the coupling of the fermions to the vectors
obtained from the solutions.
 This information  can be
summarized through the  Lagrangian density
\begin{eqnarray}
\label{Lagrangian}
{\mathcal{L}}=\frac{g}{2\sqrt{2}}[\nu^\dagger(1-I
\gamma_5)\gamma_0\gamma^\mu    u  W^+_\mu +hc]-\\ \nonumber
\frac{e}{2} [tan(\theta_W)(2l_R^\dagger\gamma_0\gamma^\mu
l_R+\nu^\dagger\gamma_0\gamma^\mu\nu+l_L^\dagger
\gamma_0\gamma^\mu l_L)+\\ \nonumber
cot(\theta_W)(\nu^\dagger\gamma_0\gamma^\mu\nu-l_L^\dagger
\gamma_0\gamma^\mu l_L)]Z_\mu- e u^\dagger \gamma_0\gamma^\mu
uA_\mu,  \nonumber
\end{eqnarray}
where $\nu$, $l_L$ are given on Table \ref{doubletend},  $l_R$
are given on Table   \ref{singletend} and $u$ are given on Table
\ref{eleMassend}. The electric charge is given by
$e=gg^\prime/{\sqrt{{g^\prime}^2+g^2}}$.

 In addition,  the vertices  give information on the coupling constants  $g^\prime$ and $g$, which
cannot be extracted from eqs. as \ref{Weinangle} or \ref{MassZ}.
Information on these can be obtained by calculating the overlap
of the vectors with the corresponding fermion currents, which are
given explicitly in eq. \ref{Lagrangian}. This can be done more
realistically
 by considering the $7+1$ dimensional Clifford algebra where vector massless solutions
become possible. The coupling constant $g$ can be obtained from
the coupling of   the massive charged vectors  $W_{M\mu}^+$ in
eq.  \ref{WsMass}  and the charged current, obtained from the
neutrino and charged massive lepton wave functions, represented
by the first term of eq. \ref{Lagrangian}. It is
\begin{eqnarray}
\label{g} g=1/\sqrt{2}\approx .707\ .
\end{eqnarray}
 %CHECK if HIGGS Terms don't overpredict. (THEY Mix left-handed and right
%handed terms as mass terms).
The coupling $g^\prime$ is deduced from the second term in eq.
\ref{Lagrangian} to be
\begin{eqnarray}
\label{gprime} g^\prime=1/\sqrt{6}\approx .408\  .
\end{eqnarray}
It is a consistency feature of the theory that these values agree
with Weinberg's angle in  eq. \ref{sinWeinangle}. (At 5+1
dimensions the also consistent  values $g=1$,
$g^\prime=1/\sqrt{3}\approx .577$ are obtained).
 In addition, these values are to be compared with the experimentally measured ones
at energies  of the mass  of the $W$ particle, which is where the
breakdown of the $SU(2)_L\times U(1)_Y$ symmetry occurs. These
are $g_{exp}^\prime\approx .35$, $g_{exp}\approx .65$, and
$sin^2(\theta_{Wexp})\approx .23$\ .

\section{Summary   and conclusions}
In this work  we have  departed from a generalized  Dirac equation whose
solutions, with  the rules we have postulated to interpret them,
exhibit some
similarity to    quantum fields.
 We have first studied
these  in the framework of the 3+1 Clifford matrices.
 They   comprise
non-axial and axial vectors, spin zero particles, antisymmetric tensors,
and, under a choice of the Lorentz generators, even spin 1/2 particles of a given chirality.
We have  investigated a   gauge symmetry of the equation.  A comparison among the different
solutions is possible with the application of a generalized point product within the quantum
mechanical framework of the equation.   Through
it the transition amplitude of a vector field and two fermions is a
 vertex, and hence, it is interpreted as an interaction. The
coupling constant
is  then determined.

We have also investigated the simplest  generalization  of the equation which is in
the context of a  5+1 dimensional Clifford algebra.
By focusing on the 3+1 underlying structure we have obtained an $SU(2)_L\times U(1)$
 symmetry. We get a
boson  and fermion set of  solutions for the massless case.
The addition of  a mass term to the equation  implies
the breaking of the symmetry to a $U(1)_Q$, which can be interpreted as the
gauge symmetry defining  the electromagnetic
interaction.  We  also  obtain the field
solutions, their spectrum, and some of the couplings
among them. We have shown they exhibit a close similarity to the particles
and coupling constants  in the  $SU(2)_L\times U(1)$  sector of the standard model
at symmetry breaking.

%Lorentz-gauge
The main result  in this work has
been to derive gauge interactions and the particle spectrum
 from an extended spin space,
 some of whose components transform
under  the usual Lorentz generators in
 3+1 space-time.
The gauge forces emerge as excitation  determined by the symmetries permitted
by the Clifford algebra in which  the 3+1 subalgebra is embedded.
Thus, we find a  relation between    gauge and space-time symmetries.
In    the simplest Clifford algebra containing the 3+1 subalgebra,
we have found a symmetry as large as $U(2)_L\times U(2)_R$  and  we have shown we have only two choices
for a model with a $SU(2)_L\times  U(1)$ which  contains both fermions and bosons.
%CHECK
The $SU(2)_L\times U(1)_L$ symmetry  group
 is consequently derived rather than being imposed. It is noteworthy
that the  chiral nature of the $SU(2)$ gauge interaction is predicted.
 The  formalism also
 predicts gauge vector carriers which are as well
generators lying in the adjoint  representation of the group.

%interactions
In general,  a  field theory
is determined by the couplings among fields, which are defined at tree level.
The power of field theory in describing nature stems from the possibility of
using this  simple description in perturbation theory to account
for  more complex behavior by  considering  repeated interactions.
The values of the coupling constants are
arbitrary and must be fixed by experiment.
%THIS SOUNDS TOO CORNY. PROPOSE ALTERNATIVE.
In our case, the very
nature of the fermion and boson solutions
defines the coupling  at tree level.  In fact, our theory determines the type of fields involved and
the normalization restriction fixes the   values of  the
coupling constants.
It is the compositeness feature of the solutions, the fact that some may be constructed
from the product of others, that determines their interaction.
For example, the form of  the spin 1/2 particle pair coupling
to vectors and scalar particles is restrained
by the symmetries of the theory.
%ALSO HERE
Thus, the restrictiveness in the choice of  the representations in our theory  constitutes  also its asset.

%fermion,
 In the model
described in eq. \ref{geneqIII}
 we have obtained   leptons with the correct gauge quantum
numbers corresponding to  a left-handed doublet of $SU(2)$ (that is,
in the fundamental representation) with
hypercharge $Y=-1$, and a singlet with $Y=-2$, which can be
interpreted as  massless neutrinos and  charged leptons. These fields
 appear in doublets characterized by a conserved quantum
number which does not affect interactions with vector bosons and which we have therefore
 associated with flavor. The flavor doublets  are   a consequence of using
a Hamiltonian which allows for a certain matrix solution space, although the size of the flavor
multiplet can change in higher dimensional models.
This may constitute
  a clue on the puzzle of generations. Furthermore, the fermions have a  conserved lepton
number.
 Thus the  fermions obtained
could be
identified with any pairs of the particle set $e$,  $\nu_e$;  $\mu$, $\nu_\mu$; and $\tau$, $\nu_\tau$.

 %Higgs spectrum and massive terms

 We  have also  obtained a spinless boson doublet with  $Y=-1$ which can  be identified with a Higgs
particle.
It is interesting   that this boson appears here as part of the solution representations and  not
put in by hand. We obtain that introducing a mass term into the equation, as represented by eq. \ref{eqIIImass}, implies an additional
interaction of the scalar particle which gives masses to some of the fields. We have
shown the $SU(2)_L\times U(1)_Y$ symmetry is broken
to a $U(1)_Q$ symmetry.
 This procedure goes further than the
SM where the Higgs mechanism is a mathematical
device to create massive terms,  and which  requires explicitly that $U(1)_Q$ remain unbroken.
In our case the presence  of a mass term implies
it is $U(1)_Q$ the unbroken  gauge symmetry in the real world.

We have obtained masses for  the vector bosons different  to
those in the SM. We have ascribed this difference to the fact
that the $6-d$ model does not allow for massless vector
solutions, which is permitted in the next Clifford algebra at 7+1
dimensions. It is encouraging that the values obtained then for
the coupling constants are within $7\%-15\%$ of their values in
the SM at electroweak breakdown. The accordance of the values of
the coupling constants, vertices  and particles described in the
theory is further fortified  by the fact that other  reducible
representations will reproduce only some aspects while others
will change. For example, the trace, which fixes the interaction,
is representation-dependent.

%gauge problem
There remain several aspects to be studied. The argument leading to the quantization condition
in eq. \ref{quanLorentz}
implies the present equations
carry on with them an implied gauge-fixing. While we have shown in detail
the extent to which this is true in the abelain case, we still have to
prove this for the non-abelian case. Analogy with the non-abelian QFT description
requires ghosts to satisfy unitarity. These are scalar objects lying in the adjoint
representation which do not appear as physical particles. We speculate
 spurious degrees of freedom as  the antisymmetric  $n$ and $v$  could conform such a counterpart in extended theories.
We have  a different characterization of the spinless  and  $Z_\mu$, $W_\mu$ fields
from  the  corresponding ones  in the SM in relation
to their discrete transformation properties, since we get
different weights for their scalar or psudoscalar, and $V$, $A$ contributions, respectively.
As the  $Z_\mu$, $W_\mu$ particles interact only weakly
and this characteristics fit into their interaction scheme
this aspect is,   however,  difficult to test.
  Furthermore, the interactions among vectors need further study.
We attribute  the mass of the lepton to be of the same magnitude
as for the $W_\mu$ to the unified approach we use.  The
possibility of this theory providing information on the
corrections to lepton masses  is under investigation.

%quantization
Finally, we have heuristically obtained fields  which
reproduce properties of  quantized operators in a quantum field theory. The implication is that
quantization is not derived  as a condition
on the fields but as a consequence of the definitions of the equations.
This points out at a closer relation to quantization which should
be researched with more detail in the future. Causality and unitarity requirements
demand   more  study on
related  aspects as propagators, commutation relations,
 and   how to include radiative corrections.
%What is needed to do CONNECT WITH ABOVE.

The presence of   bosons and fermions solutions  became possible from
the use of bi-spinors as solution space. Further extensions with the use  of more spin indices will allow for  a description of spin 3/2 and spin 2 objects;
 this may point to a  connection to gravity. This possibility can be used in turn
to propose  a new   interpretation of the wave function, with the implication of a
a closer connection of it  to space-time. The development of this idea is done from another standpoint
elsewhere (Besprosvany, 2000).
%\cite{BesproWonder}

The similarity in the representation of the
fields in this formalism and the operators which carry out a Lorentz transformation for the spin parts
could imply a possible connection between these two.
Thus,
a  Lorentz transformation could be considered not  independent of the fields
needed to perform it physically.
On the other hand,  just as the choice  of gauge interaction is restricted  by the Clifford algebras,
we also find that the  interactions restrain
the possible type of  space-time symmetry.  In this way we obtain a possible clue on the origin
of the number of dimensions of space-time. Thus, although its (3,1)
  structure is not predicted it is among  the few which are  consistent with a $SU(2)_L\times U(1)$ symmetry,
in the 5+1 dimensional Clifford algebra.

The unified treatment of space-time and gauge symmetries proposed
here has proven fruitful. The formalism presented  has the
quality of  giving information on a set of  representation
solutions and their interactions
 by literally restricting them.
Their agreement with aspects of the standard model  makes the theory a plausible alternative,
all the more that it  assumes a rather conventional
 relativistic quantum mechanical framework,   of proven  simplicity and universality.
Information on additional aspects of the standard model may be found with
the application of the theory in extended spaces, making certainly
worth its further study.

\noindent{\bf Acknowledgements}

The author acknowledges support from  DGAPA-UNAM through
project IN127298 and CONACYT through project 3275-PE.

\noindent{\bf Appendix}

We set here the conventions for the Clifford algebras used  in this work and we present explicitly the matrices
generating it.

In 4-$d$ we use the metric
\begin{eqnarray}\label{metric}
g_{\mu\nu}=
 \left( \begin{array}{cccc}
1 & 0 & 0 &  0 \\
0 & -1 & 0  & 0 \\
0 & 0 & -1  & 0 \\
0 & 0 & 0  & -1
\end{array} \right).
\end{eqnarray}

The $4\times 4$ matrices in the paper are in the Dirac representation, and in order to define them we use
the Pauli matrices $\sigma_i$, $i=1,2,3$ and the $2\times2$ unit matrix $1_2$:
\begin{eqnarray}\label{gamma0}
 \gamma_0=\sigma_3\otimes 1_2=\left( \begin{array}{cc}
1_2  & 0   \\
  0  & -1_2
\end{array} \right),
\end{eqnarray}
so the vector ${\boldmathgamma}=(\gamma^1,\gamma^2,\gamma^3)$ is given by
\begin{eqnarray}\label{gammavec}
 \boldmathgamma=i\sigma_2\otimes \boldmathsigma=
\left( \begin{array}{cc}
0  & \boldmathsigma\\
  -\boldmathsigma& 0
\end{array} \right).
\end{eqnarray}
From here all other matrices can be defined. For example,
\begin{eqnarray}\label{gamma5}
 \gamma_5=-i\gamma_0\gamma_1\gamma_2\gamma_3=\sigma_1\otimes 1_2=
\left( \begin{array}{cc}
0  & 1_2   \\
  1_2  & 0
\end{array} \right).
\end{eqnarray}

For the 6-$d$ Clifford algebra we use
\begin{eqnarray}\label{metric6}
g_{\mu\nu}=
 \left( \begin{array}{cccccc}
1 & 0 & 0 &  0 & 0 & 0 \\
0 & -1 & 0  & 0 & 0 & 0 \\
0 & 0 & -1  & 0 & 0 & 0 \\
0 & 0 & 0  & -1& 0 & 0 \\
0 & 0 & 0  & 0& -1 & 0 \\
0 & 0 & 0  & 0& 0 & -1
\end{array} \right).
\end{eqnarray}

The definitions  leading to eqs. \ref {sustitution} and  \ref{sustitutionsca}
 imply the 4-$d$ vectors subset of $8\times 8$  matrices are given explicitly by
\begin{eqnarray}\label{gammamup}
 \gamma_\mu^\prime=1_2\otimes\gamma_\mu=
\left( \begin{array}{cc}
\gamma_\mu & 0\\
  0  & \gamma_\mu
\end{array} \right),\ \  \mu=0,1,3,
\end{eqnarray}

\begin{eqnarray}\label{gamma2p}
 \gamma_2^\prime=\sigma_1\otimes\gamma_2=
\left( \begin{array}{cc}
0  & \gamma_2\\
  \gamma_2 & 0
\end{array} \right),
\end{eqnarray}
and the 4-$d$  scalars by
\begin{eqnarray}\label{sca4}
1_8=1_2\otimes1_4=
\left( \begin{array}{cc}
1_4 & 0\\
  0  &1_4
\end{array} \right),
\end{eqnarray}
\begin{eqnarray}\label{sca1}
I=\sigma_1\otimes 1_4=\left( \begin{array}{cc}
0 & 1_4\\
  1_4  & 0
\end{array} \right),
\end{eqnarray}
\begin{eqnarray}\label{sca2}
i\gamma_5^\prime=i J \gamma_2=i\sigma_2\otimes\gamma_2=
\left( \begin{array}{cc}
0&  \gamma_2\\
  -\gamma_2  & 0 \end{array} \right),
\end{eqnarray}
\begin{eqnarray}\label{sca3}
i\gamma_6^\prime=i K \gamma_2=i\sigma_3\otimes\gamma_2=
\left( \begin{array}{cc}
i\gamma_2 & 0\\
  0  & -i\gamma_2
\end{array} \right).
\end{eqnarray}
All $8\times 8$ matrices can  be generated by products of these matrices.
We use a notation for  which the $\gamma_\mu^\prime$ matrices  are written in
terms of the $\gamma_\mu$ matrices and   from eq. \ref{sustitution} onwards the latter are assumed
to be $8\times 8$ matrices.

\setcounter{equation}{0}

%\hskip


\begin{thebibliography}{99}

\bibitem{Bargmann} V.  Bargmann and E. P. Wigner, Proc. Nat. Acad. Sci. (USA) 34 (1948) 211.

%\bibitem {Bell} J. S. Bell, Speakable and Unspeakable in Quantum Mechanics (Cambridge
%University Press, Cambridge, 1987).

\bibitem{BesproWonder} J. Besprosvany, Submitted (2000).



\bibitem{Dirac} P. A. M. Dirac,   The Principles of Quantum Mechanics, Fourth Edition, Claredon Press, Oxford, 1958.


%\bibitem {EPR} A. Einstein, B. Podolsky,  and N. Rosen, Phys. Rev. 47 (1935) 777.
 %Is this the right order of the names?

\bibitem {unification} H.  Georgi and S. L. Glashow, Phys. Rev. Lett. 32 (1974) 438.

\bibitem {Glashow} S. Glashow, Nucl. Phys. 22 (1961) 579.


\bibitem {Yang} T. D. Lee and C. N. Yang, Phys. Rev. 104 (1956) 254.


\bibitem{Salam} A. Salam in Elementary
Particle Theory, (ed. W. Svartholm, Almquist and Wiskell, Stockholm, 1968).

\bibitem {Weinberg} S. Weinberg, Phys. Rev. Lett. 19 (1967) 1264.



\end{thebibliography}
\end{document}